\documentclass[useAMS,usenatbib]{mn2e}
\usepackage{graphicx,rotate,url,mathptmx,lscape,times,amsmath,color,multirow,amssymb}
\usepackage{aas_macros}
\font\manual=manfnt at 7pt \def\dbend{\hbox{\raise0.9ex\hbox{\manual\char127\hspace{0.6em}}}}

\newcounter{INTERNALionstage}

\def\gtsim{\mathrel{\hbox{\rlap{\hbox{\lower4pt\hbox{$\sim$}}}\hbox{$>$}}}}
\def\lesssim{\mathrel{\hbox{\rlap{\hbox{\lower4pt\hbox{$\sim$}}}\hbox{$<$}}}}

\def\ga{{\rm\thinspace gauss}}

\def\la{\mbox{{\rm L}$\alpha$}}

\def\h0{\mbox{{\rm H}$^0$}}

\DeclareMathAlphabet{\vib}{OML}{cmm}{m}{it}

\title[H-, He-like recombination spectra III]{H-, He-like recombination
  spectra III: $n$-changing collisions in highly-excited Rydberg states and
  their impact on the radio, IR and optical recombination lines}

\author[F. Guzm\'an et al.]
       {\parbox[]{6.0in}
         {F. Guzm\'an$^1$, M. Chatzikos$^1$, P. A. M. van Hoof$^3$,
           Dana S. Balser$^2$, M. Dehghanian$^1$, N. R. Badnell$^4$ and
           G.J. Ferland$^1$. \\
        \footnotesize
        $^1$Department of Physics and Astronomy, University of Kentucky,
        Lexington, KY 40506, USA.\\
        $^2$NRAO, Charlottesville, VA 22903-2475.\\
        $^3$Royal Observatory of Belgium, Ringlaan 3, 1180 Brussels, Belgium.\\
        $^4$Department of Physics, University of Strathclyde, Glasgow G4 0NG, UK.}
}

\date{In preparation \today }

\pagerange{\pageref{firstpage}--\pageref{lastpage}}
\pubyear{}

\begin{document}

\maketitle

\label{firstpage}

\begin{abstract}

\noindent
At intermediate to high densities, electron (de-)excitation collisions are the
dominant process for populating or depopulating high Rydberg states. In
particular, the accurate knowledge of the energy changing ($n$-changing) 
collisional rates is determinant for predicting the radio recombination spectra 
of gaseous nebula. The different datasets present in the literature come either 
from impact parameter calculations or semi-empirical fits and the rate
coefficients agree within a factor of two. We show in this paper that these
uncertainties cause errors lower than 5\% in the emission of radio
recombination lines (RRL) of most ionized plasmas of typical nebulae. However,
in special circumstances where the transitions between Rydberg levels are
amplified by maser effects, the errors can increase up to 20\%. We present
simulations of the optical depth and H$n\alpha$ line emission of Active
Galactic Nuclei (AGN) Broad Line Regions (BLRs) and the Orion Nebula Blister to
showcase our findings. 
\end{abstract}


\section{Introduction}
\label{sec:intro}

Radio observations are an important tool in astronomy as the spectrum is not
affected by dust extinction. Radio recombination lines (RRL) coming from
highly-excited Rydberg levels are used to determine the temperatures and 
densities of gaseous nebulae \citep[see for example][]{AGN3}, the metallicity 
structure of the Galaxy \citep{Balser2015,Balser2011}, to survey the H~II 
regions of our Galaxy \citep{Anderson2018} and determine their structure
\citep{Anderson2015,Poppi2007}, to probe extragalactic Active Galactic Nuclei 
(AGN) and startburst galaxies \citep{Scoville2013}, to study the 
Photo-Dissociation Regions \citep{Luisi2017}, and to study the diffuse ionized 
gas in H~II regions and the diffuse clouds of the cold neutral medium 
\citep{Morabito2014}. 

In particular, hydrogen and helium RRL can be used to obtain the electron
temperature $T_e$. If the levels involved in the line transition are close to
local thermodynamic equilibrium (LTE), $T_e$ can be obtained directly from the
ratio of the line brightness temperature to the continuum brightness
temperature \citep{Rohlfsandwilson2000}. However, non-LTE effects can be
important for RRL \citep{Brown1978}. Likewise, stimulated emission (maser
effects) and pressure broadening effects can alter the equilibrium of lines
\citep{Shaver1980}. In these cases, modeling of the level populations using
collisional radiative (CR) modeling
\citep{Dupreeandgoldberg1970, Brocklehurstandseaton1972,Burgess1976} is needed
to obtain the correct line intensities. CR modeling is subject to the
availability and accuracy of the fundamental atomic data describing the
processes between the different particles of the emitting gas. The highest
uncertainties come from the collisional (de-)excitation, which can be divided
into two types of interaction: energy changing and angular momentum changing.
In the previous two papers of this series, devoted to investigate and assess
these collisional data, which have a high impact on astronomical observations
\citep[][hereafter Paper I and Paper II]{Guzman.I.2016,Guzman.II.2017}, we
showed how the disparities in the theoretical values of $l$-changing collisions
cross sections can affect the emissivities of the optical recombination lines.
$l$-changing collisions are mainly produced by ion impact and alter the angular
momentum of the target electron by Stark mixing without changing its energy 
\citep{PengellySeaton1964}. As shown in Paper I and Paper II, at high Rydberg 
states, $l$-changing collisions are so fast that they populate the $l$
subshells statistically, and differences of one order of magnitude between the
values of the rate coefficients are not important. However, as noted by
\cite{Salgado.I.2017}, the low densities of the cold neutral medium make the 
Stark effect interactions of $l$-changing collisions important at the high 
levels involved in RRL.

This paper is the third of the series and is focused on electron impact
excitations changing the principal quantum number $n$ of the target electron.
These energy changing electron collisions redistribute the electronic
population of the high Rydberg levels of ions and atoms towards LTE. However,
many of the Rydberg states will remain in non-LTE. These effects will mainly
depend on the accuracy of the electron-impact excitation rate coefficients,
which can significantly influence the derived opacities and the temperature
diagnostics obtained with RRL. Unfortunately there are no precise quantum
calculations for collisional excitation of high Rydberg levels. R-matrix
calculations for $n$-changing collisional data for hydrogen- and helium-like
(H-like and He-like) iso-sequences reach up to $n=5$
\citep{Anderson2000,Bray2000,Ballance2006,Ralchenko2008,Griffinandballance2009}.
Analytic formulas using the Born approximation \citep{Lebedevandbeigman1998} and
fitting formulas from theoretical and experimental data 
\citep{Fujimoto1978,Vriens1980} have been used for transitions with higher $n$.

In this paper, we analyze the impact of the different approaches for
$n$-changing electron-impact excitation cross sections on the Rydberg
populations of H- and He-like ions and on the RRL. Our main goal is to
identify the best data and quantify how the determination of the physical
parameters derived from the observed RRL can be compromised by the uncertainty
in the $n$-changing collisions. In Section \ref{sec:datasets}, the different
approximations are discussed and compared. In Section \ref{sec:RRL}, we
discuss how the different values of $n$-changing rate coefficients translate
to the populations densities and the predicted RRL intensities and opacities.
In that Section, we analyze a simple slab of irradiated gas using the spectral
simulation code Cloudy, last described by \citet{CloudyReview}. The
uncertainties on the excitation rates at high Rydberg levels could affect the
lower levels by cascading decay, and thus compete with the errors from the
collisional data for low lying levels. We discuss these effects on the
uncertainty of the predicted emissivities of optical recombination lines 
in Section \ref{sec:optlines}. In Section \ref{sec:agn} and \ref{sec:orion} we
apply our analysis to models of the Broad Line Regions (BLRs) in AGN and the 
Orion nebula emission, respectively. Finally, we summarize our results in the 
conclusions. 

\section{Rydberg collisional excitation theories}
\label{sec:datasets}

Electron-impact excitation between high $n$ Rydberg ($n\geq 5$) levels have
received considerably less attention than between low-lying $n$-shells due
to a number of reasons. First, the low-lying transitions in the optical part of
the spectrum are much brighter than the radio lines and less affected by
pressure broadening. Line ratios are routinely used in the optical to obtain
temperatures and densities. Second, dust absorption and scattering peaks at the
mid infrared and is still big at the far infrared \citep{AGN3} making 
intermediate-high Rydberg levels transitions unobservable in dusty 
environments. Third, the numerical and computational limitations of the 
theoretical methods prevent accurate quantum calculations for transitions to 
higher levels. For example, in most R-matrix calculations, including higher 
levels requires the appropriate inclusion of pseudostates to correctly cover 
excitation in the close coupling terms. The number of pseudostates needed will 
grow with the inclusion of more levels. \citet{Ballance2006} show how 
including the coupling to pseudostates modified their results between 
40\% -- 60\%. Finally, in very low density environments, such as in the 
interstellar medium (ISM), low levels recombine faster and the Rydberg levels 
eventually decay radiatively to lower levels. At high densities, collisions 
are so fast that the levels are expected to be in LTE independently of the 
value of the collisional transition probability, so the actual theory is 
unimportant. However, there is a wide number of astrophysical circumstances
where the Rydberg levels can be in non-LTE and their populations driven by the
rates of electron impact collisions (e.g. see Sections \ref{sec:RRL} and
\ref{sec:models}).

\begin{figure}
  \begin{center}
    \includegraphics[width=0.5\textwidth,clip]{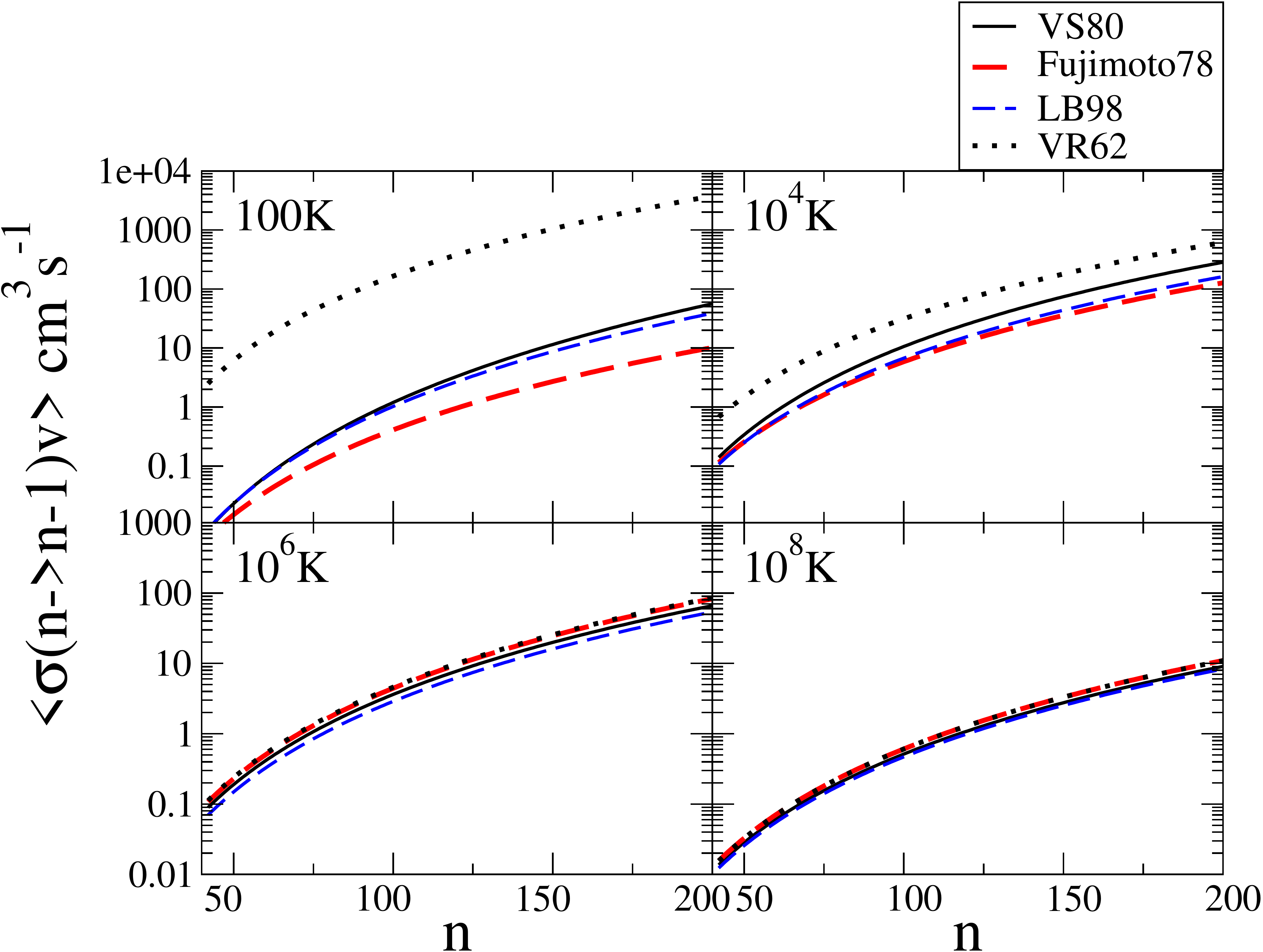}
    \caption{\label{f:nnm1} Effective rate coefficients for $n\to n-1$
      electron-impact transitions as a function of $n$ for different electron
      temperatures (indicated at the top left of each panel). VS80:
      \citet{Vriens1980}; LB98: \citet{Lebedevandbeigman1998}; Fujimoto78:
      \citet{Fujimoto1978}; VR62: \citet{VanRegemorter1962}.}
  \end{center}
\end{figure}

\subsection{Collisional excitation datasets}
Maxwell-averaged collisional effective rate coefficients for the dominant
$n\to n-1$ de-excitations, where $n$ is the principal quantum number of Rydberg
states, are plotted in Figures \ref{f:nnm1} and \ref{f:nnm1vst} for neutral
atoms and for $T_e=100-10^8$K as a function of the upper level main quantum
number $n$ and the electron temperature respectively. While the curves have a
similar shape (with the exception of the VR62 set explained later in this
section, they differ from 30\% at high temperatures to 60\% at
$T_e\sim10^4-10^5$K. In this paper we investigate whether these uncertainties
could impact predictions of astrophysical interest.

\subsubsection{VR62}

\begin{figure}
  \begin{center}
    \includegraphics[width=0.5\textwidth,clip]{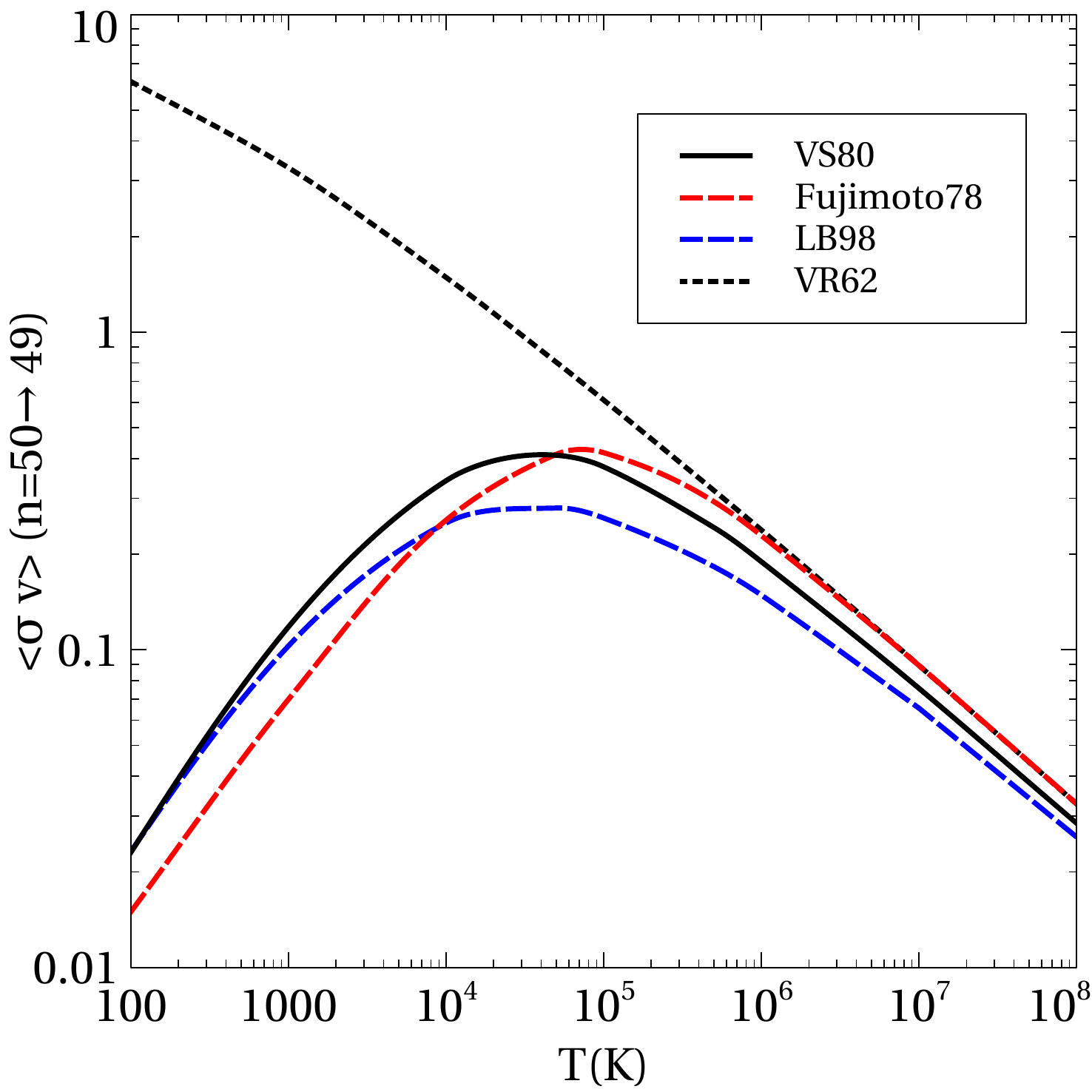}
    \caption{\label{f:nnm1vst} Effective rate coefficients for $n\to n-1$
      electron impact de-excitation for $n=50-49$ as a function of the
      electron temperature. Legend labels as in Fig. \ref{f:nnm1}.}
  \end{center}
\end{figure}

\citet{VanRegemorter1962} obtained new values for the experimentally averaged
Gaunt factor $\bar{g}$ approximation proposed by \citet{Seaton1959a}. The
effective rates for de-excitation proposed by this model are:
\begin{equation}
  q_{n^\prime\to n} = \frac{20.60\lambda^3A_{nn^\prime}}{\sqrt{T_e}} P(\epsilon) 
\quad\text{cm}^3\text{s}^{-1}\,,
  \label{eq:vrgm}
\end{equation}
\noindent where $T_e$ is the electron temperature in K,
$\lambda = \frac{hc}{\Delta E_{nn^\prime}}$(cm), with $h$ the Planck constant
and $c$ the speed of light, $\Delta E_{nn^\prime}$ is the $n^\prime\to n$
transition energy, $A_{nn^\prime}$ is the Einstein coefficient,
$\epsilon = \frac{\Delta E_{nn^\prime}}{kT_e}$, and
\begin{equation}
  P(\epsilon)=\int_0^\infty\bar{g}(x)\exp(-\epsilon x^2)\text{d}(\epsilon x^2)\,,
  \label{eq:pe}
\end{equation}
\noindent with $x=\sqrt{\frac{m_ev^2}{\Delta E_{nn^\prime}}}$
$\left( \text{so} \quad \epsilon x^2 = \frac{m_ev^2}{kT_e}\right)$, $m_e$ the
rest mass of the electron, $\nu$ the frequency of the transition
$n\to n^\prime$, and $\bar{g}(x)$ a Gaunt factor
\citep[see e.g.][]{Seaton1959a} obtained from fitting experimental values.
\citet{VanRegemorter1962} provides values for the function $P(\epsilon)$ of
equation (\ref{eq:pe}) in his table 2. He used the Born approximation
together with experimental results to obtain the functions $\bar{g}(x)$. We
have interpolated these values to calculate the dotted curve in Figs.
\ref{f:nnm1} and \ref{f:nnm1vst}. It is important to note that
\citet{VanRegemorter1962} fitted his Gaunt factors to transitions between low
lying levels of atoms and ions (see figure 2 of his paper) relevant for UV
observations of the solar corona. The VR62 results greatly overestimate later
calculations for high Rydberg levels at low and intermediate temperatures and
should not be used at low temperatures. We have included them here for
comparison purposes. These results are in better agreement with the
semi-empirical formulas of Fujimoto78 at high energies, where the correct
asymptotic behavior of $T^{-1/2}$ is ensured. This is shown in Figure
\ref{f:nnm1vst} (more details in Section \ref{sec:F78}).

\subsubsection{Fujimoto78}
\label{sec:F78}

\citet{Fujimoto1978}, in an internal report, integrated experimental and
theoretical cross section values fitted by \citet{Johnsonandhinnov1969} to a
fitting formula for rate coefficients. These authors used measurements of
population densities in an experimental plasma at the C stellarator
\citep{Sinclair1965} and compared with the solution of the CR balance
equations for each level to deduce the excitation cross sections. The
experiments were done for a range of electron temperatures between 0.04 eV to
1 eV (464K to 11604K) and electron densities between $10^{11}\text{cm}^{-3}$
and $3\times10^{13}\text{cm}^{-3}$. Asymptotic forms from the impact parameter 
approximation \citep{Seaton1962} were used for higher temperatures. The fits
of \citet{Fujimoto1978} have been recommended by \citet{Ralchenko2008}
for helium transitions with upper and lower $n>5$. The de-excitation effective
coefficients are:
\begin{align}
  q_{n^\prime\to n}&=e^{-\epsilon}K\left[\frac{\ln(1+\delta)-
    e^{(1+\delta)\epsilon}E_1\left(-(1+\delta)\epsilon\right)}{\epsilon}
  \right] \nonumber\\
  &\quad -\frac{e^{-2\beta(f_{nn^\prime}\text{I}_{\text{H}}/
      \Delta E_{nn^\prime})^{-\gamma}}}{\beta(f_{nn^\prime}\text{I}_{\text{H}}/
    \Delta E_{nn^\prime})^{-\gamma}+\epsilon} \nonumber \\
  &\quad \times \Big\{\ln(1+\delta)+e^{(1+\delta)\left[\beta(f_{nn^\prime}
\text{I}_{\text{H}}/\Delta E_{nn^\prime})^{-\gamma}+\epsilon\right]} \nonumber\\
  &\quad  E_1\left((1+\delta)\left[\beta(f_{nn^\prime}\text{I}_{\text{H}}/
    \Delta E_{nn^\prime})^{-\gamma}+
      \epsilon\right]\right)
    \Big\}\,,
    \label{eq:fuji}
\end{align}
\noindent where
\begin{equation}
  K = 8\left(\frac{\text{I}_{\text{H}}}{kT_e}\right)^{3/2}\sqrt{\pi}\alpha ca_0^2 
  f_{nn^\prime} 
  \label{eq:kfuji}\,.
\end{equation}
In equations (\ref{eq:fuji}) and (\ref{eq:kfuji}) $\text{I}_{\text{H}}$ is the
Rydberg energy, $f_{nn^\prime}$ is the oscillator strength of the $n-n^\prime$
transition, $\alpha$ is the fine structure constant, $a_0$ is the Bohr radius,
$\epsilon= \frac{\Delta E_{nn^\prime}}{kT_e}$ as in equations (\ref{eq:vrgm})
and (\ref{eq:pe}), $E_1(x)$ is the first exponential integral
\citep[see][]{abramowitz1965} and $\beta=0.5$, $\gamma=0.7$ and $\delta=0.2$
are fitting constants. Since \citet{Johnsonandhinnov1969} use results from the
impact parameter approximation \citep{Seaton1962}, and these results have been
also used by \citet{VanRegemorter1962} for his $\bar{g}$ approach, it is not
surprising that Fujimoto  and Van Regemorter datasets coincide in the limit of
the high electron temperatures. That can be seen in Fig. \ref{f:nnm1}, and
specially in Fig. \ref{f:nnm1vst}. \citet{Johnsonandhinnov1969} report high
uncertainties in their cross sections for high $n$ because these levels are
closer to LTE, and the value of the cross sections does not have an important
influence in the populations.

\subsubsection {VS80}

\citet{Vriens1980} fit experimental results for intermediate-high energies
using several theoretical and experimental works available at the time. Their
excitation coefficients are mainly based on the semi-empirical formulas of
\citet{Gee1976}. These authors recommend their results for
$10^6\text{K}/n^2 < T_e \ll 3\times 10^9\text{K}$. \cite{Vriens1980} overcame
this limitation by using their own extrapolation at lower energies,
simultaneously fitting  experimental and theoretical results, including
\citet{Percival1976} semi-empirical formulae, \citet{Johnson1972} Bethe-Born
approximation and available experimental data
\citep[see references within][]{Vriens1980}. The final de-excitation rate
coefficients are given by:
\begin{align}
  q_{n^\prime\to n} = & 4.33\times10^{-8}\frac{g_n}{g_{n^\prime}} 
  \frac{
\left(\frac{kT_e}{\text{I}_{\text{H}}}
\right)^{1/2}}
{\frac{kT_e}{\text{I}_\text{H}} + \Gamma_{nn^\prime}}\nonumber \\
  & \times\left[\mathcal{A}_{nn^\prime}\ln\left(\frac{0.3kT_e}{\text{I}_\text{H}}+
    \Delta_{nn^\prime}\right) + \mathcal{B}_{n^\prime n}\right]
  \text{cm}^3\text{s}^{-1}\,\text{,}
  \label{eq:vs80}
\end{align}
\noindent where $g_n$ and $g_{n^\prime}$ are the statistical weights of the
states $n$ and $n^\prime$, and
\begin{subequations}
\begin{align}
  \Gamma_{nn^\prime} & = \ln\left(1+\frac{n^3kT_e}{\text{I}_{H}}\right)
  \left[3+11\left(\frac{\left|n^\prime-n\right|}{n}\right)^2 \right] \nonumber \\
    & \quad\times \Big(6 +1.6n^\prime\left|n^\prime-n\right|+ 
  0.3\left|n^\prime-n\right|^{-2} \nonumber \\
   & \quad+0.8 n^{3/2}\left|n^\prime-n\right|^{-1/2}
    \left|\left|n^\prime-n\right|-0.6\right|\Big)^{-1} \label{eq:gvs80}\\
    \Delta_{nn^\prime} & = \exp\left(\frac{\mathcal{B}_{nn^\prime}}
          {\mathcal{A}_{nn^\prime}}\right)+
          \frac{0.06\left|n^\prime-n\right|^2}{n^\prime n^2}
          \label{eq:dvs80} \\
     \mathcal{A}_{nn^\prime}& = \left(\frac{2\text{I}_{\text{H}}}
             {\Delta E_{nn^\prime}}\right)f_{nn^\prime} \label{eq:Avs80} \\
     \mathcal{B}_{nn^\prime} & = \frac{4\text{I}_{\text{H}}^2}{{n^\prime}^3}
             \left(\frac{1}{\Delta E_{nn^\prime}^2}+
             \frac{4E_n}{3\Delta E_{nn^\prime}^3}+
             c_n\frac{E_n^2}{\Delta E_{nn^\prime}^4}\right)
             \label{eq:Bvs80} \\
     c_n & = \frac{1.4\ln(n)}{n}-\frac{0.7}{n}-\frac{0.51}{n^2}+
              \frac{1.16}{n^3} -\frac{0.55}{n^4} \label{eq:bvs80}\,,
\end{align}
\label{eq:vs80aux}
\end{subequations}
\noindent and $E_n$ is the ionization potential of the shell $n$. The remaining 
symbols have the same meaning as in equations (\ref{eq:fuji}) and
(\ref{eq:kfuji}).

We have checked that the rate coefficients from \citet{Gee1976} agree with
\citet {Vriens1980} within 15\% over their range of validity.  Additionally,
\citet{Percival1978} modified \citet{Gee1976} coefficients to include charge
dependence. 

\subsubsection{LB98}

\citet{Lebedevandbeigman1998} use the semiclassical straight-trajectory Born
approximation to provide \emph{ab initio} excitation rate coefficients for
Rydberg atoms. The de-excitation formula for $n^\prime\to n$ is:
\begin{align}
  q_{n^\prime\to n} & = \frac{g_n}{g_{n^\prime}}
  \exp\left(\frac{\Delta E_{nn^\prime}}{kT_e}\right) q_{n\to n^\prime} \nonumber \\
  & = \frac{g_n}{g_{n^\prime}}
  2\sqrt{\pi} a_0^2 \alpha cn
  \left[\frac{n^\prime}{Z(n^\prime-n)}\right]^3\frac{f(\theta)\varphi}{\sqrt{\theta}}
  \label{eq:lebedev}\,\text{,}
\end{align}
\noindent where
\begin{subequations}
\begin{align}
  \varphi & = \frac{2{n^\prime}^2n^2}{(n^\prime+n)^4(n^\prime-n)^2}
  \left[4(n^\prime - n) -1 \right]\exp\left(\frac{E_n}{kT_e}\right)
  E_1\left(\frac{E_n}{kT_e}\right) \nonumber \\
  & + \frac{8n^3}{(n^\prime+n)^2(n^\prime-n)n^2{n^\prime}^2}
  \left(n^\prime-n-0.6\right)\left(\frac{4}{3}+n^2(n^\prime-n)\right)\nonumber\\
  & \times \left[1 - \frac{E_n}{kT_e}\exp\left(\frac{E_n}{kT_e}\right) 
  E_1\left(\frac{E_n}{kT_e}\right)\right] \label{eq:lebphi}\\
  f(\theta) & = \frac{\ln\left(1+\frac{n\theta}{Z(n^\prime-n)\sqrt{\theta}+2.5}
    \right)}{\ln\left(1+\frac{n\sqrt{\theta}}{Z(n^\prime-n)}\right)}
  \label{eq:lebftheta} \\
  \theta & = \frac{kT_e}{Z^2\text{I}_{\text{H}}} \label{eq:lebtheta}\,\text{,}
\end{align}
\label{eq:lebedevaux}
\end{subequations}

\noindent and $\alpha$ and $Z$ are the fine structure constant and the core
charge of the target ion or atom. While VS80, Fujimoto78 and VR62 formulas are
theoretical formulas fitted to experimental results, LB98 is a semiclassical
treatment (impact parameter) of the Born approximation and involves no fit to
experimental results. \citet{Lebedevandbeigman1998} note that the straight
trajectories approximation breaks at low energies and when the target ions
have high charges. Coulomb trajectories should be assumed in these cases.
Because of that, we expect an underestimation of the rates obtained with this
formula at low collision energies and $Z>1$. Fujimoto78 and VS80 formulas have
the same shape at low temperatures, but they are intended only for atoms. A
study of the collisional rates for higher charges is beyond the scope of this
paper and will be the object of future work.

\subsection{Comparison of the collisional excitation rate coefficients}

The curves in Fig. \ref{f:nnm1} are mostly parallel for all the range of 
$n\to n-1$ collisional transitions considered here. In Fig. \ref{f:nnm1vst} 
LB98, VS80 and Fujimoto78 datasets have a similar dependence with the
temperature. The $\bar{g}$ approximation used by VR62 was originally conceived
for high energies and has the correct T$^{-1/2}$ dependence at high
temperatures, failing to give the correct low energy behavior. This is a known
fact of the $\bar{g}$ approach. Fujimoto78 formula has the same high energy
dependence as VR62. However, there is reasonable doubt over the experimental
rate coefficients at temperatures above the range probed by the experiments of
\citet{Johnsonandhinnov1969}. These authors use \citep{Seaton1962} approach.
Furthermore, levels at very high $n$ might be in LTE for the high electron
densities of the experiments producing inaccurate values of the rate
coefficients. A visual inspection of figure 5 of \citet{Vriens1980}
shows that the \citet{Gee1976} values and the VS80 extrapolations do not 
differ by more than a factor $\sim2$. This is similar to the difference 
between VS80 and LB98 and Fujimoto78. \citet{Salgado.I.2017} compare the 
values of \citet{Vriens1980} with the rate coefficients given by 
\citet{Gee1976} for low temperatures obtaining differences of less than 20\%. 
LB98 {\it ab initio} rate coefficients are supported by the experimental fits 
included in the VS80 and Fujimoto78 datasets (Fig. \ref{f:nnm1vst}) even at low 
temperatures, indicating that the straight-trajectory approximation is good in 
this range at least for neutral targets. The LB98 calculations do not rely on 
extrapolations from previous experimental fittings and might be more 
accurate.

\subsubsection{$\Delta n>1$}

\begin{figure}
  \begin{center}
    \includegraphics[width=0.5\textwidth,clip]{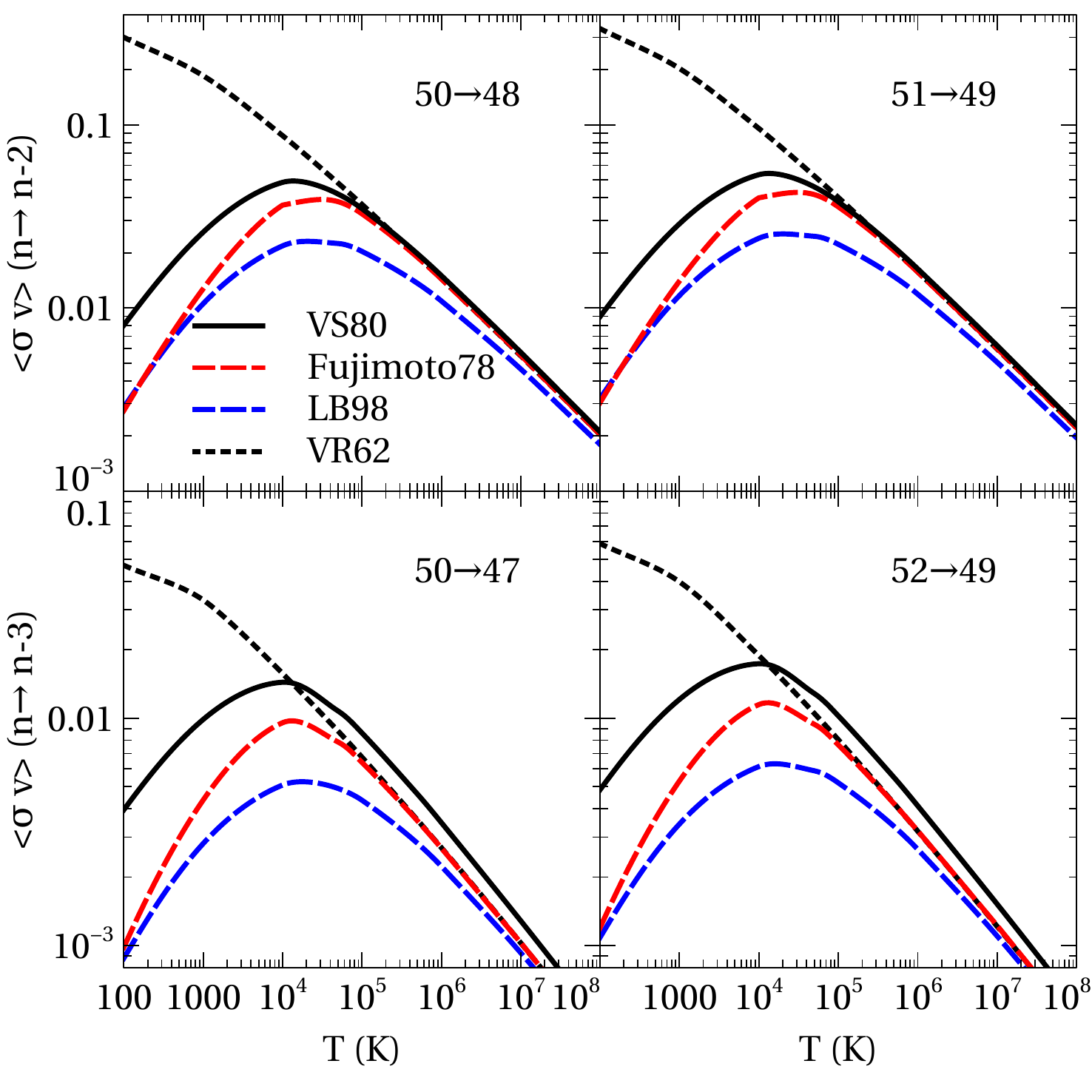}
    \caption{\label{f:nnmdn} Effective rate coefficients for $n\to n-2$ and
      $n \to n-3$ electron impact transitions as a function of temperature for 
      different $n$ indicated at the top right corner of each panel. Legend 
      labels as in Fig. \ref{f:nnm1}.}
  \end{center}
\end{figure}

Although (de-)excitation collisions with $\Delta n = 1$ are dominant,
collisions with higher $\Delta n$ can contribute to  (de-)populate high Rydberg
levels. While Fujimoto78 and VR62 depend on $\Delta n$ through
$\Delta E_{nn^\prime}$, LB98 and VS80 have a more complicated dependence. Rate
coefficients for $\Delta n<4$ are compared on the four panels of the Fig.
\ref{f:nnmdn} for collisional de-excitation transitions depopulating $n=50$ 
(left panels) and populating $n=49$ (right panels). $\Delta n>1$ collisions 
contribute to the populating and depopulating rates appreciably. For example,
at $T_e=10^4$K, $\Delta n=2$ depopulating rate coefficients for $n=50$ are
$\sim14$\% of $\Delta n=1$ collisional transitions for VS80 data and $\sim9$\%
for LB98, while $\Delta n=3$ contributes another $\sim4$\% for VS80
and $\sim2$\% to LB98. The final contribution of $1<\Delta n<4$ is then
$\sim18$\% for VS80 and $\sim11$\% for LB98 increasing the difference between
the final (de-)populating rates obtained using each dataset. $\Delta n>1$
collisions are included in the Cloudy simulations shown in this paper unless
specified otherwise.  

\section{Emission of radio recombination lines}
\label{sec:RRL}

RRL of H-like and He-like elements are produced by transitions between high
$n$-shells Rydberg levels. At high enough $n$, $l$-changing collisional
transitions overcome radiative decay and bring the $l$ subshells to
equilibrium. At this point we can consider the populations of the $l$-shells
to be statistically distributed as:
\begin{equation}
  n_{nl} = \frac{2l+1}{n^2}n_n\,.
\label{eq:Nnl}
\end{equation}
The total radiative lifetime is $\tau_{nl} = A_{nl}^{-1} \propto n^{5}$, where
$A_{nl}$ is  the total spontaneous decay rate out of level $nl$. The
collisional redistribution time can be calculated as
$\tau^{\text{coll}}_{ll^\prime} = (n_iq_{ll^\prime})^{-1}$, where $n_i$ is the
density of heavy ions\footnote{Heavy ions are more effective than electrons in
  intrashell collisions changing angular momentum \citep{PengellySeaton1964}.}
and $q_{ll^\prime}$ is the collisional effective coefficient for the
$nl\to nl^\prime$ transition. If $l\ll n$, $q_{ll^\prime} \sim n^4$
\citep{PengellySeaton1964}. The critical density where $l$-changing collisions
rates are equal to radiative transitions is defined as
$n^{\text{crit}} = 1/(\tau_{nl}q_{ll^\prime})$ and will depend on $n$ as
$\sim n^{-9}$. For $n\sim30$ in hydrogen at $T=10^4$K,
$n^{\text{crit}}\simeq 30\text{cm}^{-3}$ and for $n \sim 50$ and
$n^{\text{crit}}\simeq 0.6\text{cm}^{-3}$. Typical densities of H~II regions
are of $n_{\text{H}}\sim 10^3 - 10^4 \text{cm}^{-3}$, while diffuse atomic gas
have $n_{\text{H}}\sim 50 \text{cm}^{-3}$. We can assume that levels with
$n>50$ are statistically populated in the $l$-subshells as in equation
(\ref{eq:Nnl}). For the much lower ISM, intergalactic or intracluster
densities, this assumption is not valid and $l$-subshell populations need to
be considered. Thus, the equilibrium of the high Rydbergs is determined by the
$n$-changing collisions, where transitions $n\to n-1$ are dominant. At high
enough $n$, collisions are frequent and populations will be driven to LTE.

\subsection{Rydberg Emissivities}
\label{sec:emis}

The local emissivities for $n\to n^\prime$ transitions are given by:
\begin{equation}
  j_{nn^\prime} = n_n\frac{A_{nn^\prime}}{4\pi} h\nu\,,
  \label{eq:emiss1}
\end{equation}
\noindent where $A_{nn^\prime}$ are the $l$-averaged Einstein coefficients for
$n \to n^\prime$ transitions, and from (\ref{eq:Nnl}):
\begin{equation}
  A_{nn^\prime} = \sum_{l^\prime=0}^{l^\prime=n^\prime-1}
\sum_{l=0}^{l=n-1}
  \frac{(2l+1)}{n^2}A_{nl,n^\prime l^\prime}\,.
  \label{eq:Anl}
\end{equation}
The population can be obtained by solving the collisional-radiative system
\citep[see e.g.][]{Brocklehurst1970,AGN3}:
\begin{align}
  & n_+n_e\alpha_n + \sum_{n^\prime>n}n_{n^\prime} A_{n^\prime n} \nonumber\\
 + & \sum_{n^\prime}
  n_{n^\prime} j_\nu B_{n^\prime n} + 
  \sum_{i}\sum_{n^\prime} n^{\text{col}}_in_{n^\prime} q_{n^\prime n} \nonumber\\
  = & n_n \left[\sum_{n^\prime<n} A_{nn^\prime} + 
    \sum_{n^\prime} j_\nu B_{n n^\prime} \right. \nonumber\\
+& \left. \Gamma_n + \sum_{i}\sum_{n^\prime} n^{\text{col}}_iq_{nn^\prime} + 
\sum_{i} n^{\text{col}}_iS_n^{\text{ion}} \right ]\,.
  \label{eq:CR}
\end{align}

Here,
$\alpha_n = \alpha_n^{\text{rad}} + \alpha_n^{\text{DR}} + n_e\alpha_n^{\text{3b}}$
is the total recombination to the shell $n$, composed of radiative recombination
(including spontaneous and induced recombination), dielectronic recombination
(in case of He-like ions), and three body recombination respectively.
$A_{nn^\prime}$ and $B_{nn^\prime}$ are the Einstein spontaneous and stimulated
emission coefficients, $\Gamma_n$ is the photo-ionization integral from level
$n$, $q_{nn^\prime}$ is the collisional rate coefficient (cm$^3$s$^{-1}$) of the
transition from level $n$ to level $n^\prime$, $n_+$ is the population of the
parent ion before recombination, $n^{\text{col}}$ represent the density of
colliders responsible for this transition (electrons or heavy particles) and
$S_n^{\text{ion}}$ is the ionization rate coefficient from level $n$. In
equation (\ref{eq:CR}), radiative excitation terms are negligible for most
astrophysical systems. Radiative recombination and photo-ionization contribute
mostly to the lower lying levels, while three body recombination and
collisional ionization terms are more important for the high Rydberg levels,
approaching them to LTE.

It is common to express the population of the $n$-shell as a function of the
departure coefficients $b_n$:
\begin{equation}
  n_n = n_en_+\left(\frac{h^2}{2\pi m_ekT_e}\right)^{3/2}\frac{g_n}{2}
  e^{\frac{E_n}{kT_e}}b_n = n_n^{\text{LTE}}b_n\,,
  \label{eq:departures}
\end{equation}
\noindent where $g_n$ is the statistical weight of the shell $n$. In LTE,
$b_n=1$ and the populations $n_n$ will obey the Saha-Boltzmann law. To obtain
the departure coefficients, the CR equations (\ref{eq:CR}) for the populations
of the ions of the gas need to be solved. Substituting equation
(\ref{eq:departures}), it is possible to write the equations (\ref{eq:CR}) in
terms of the departure coefficients:
\begin{align}
  & \left(\frac{h^2}{2\pi m_ekT_e}\right)^{-3/2}\frac{\alpha_n}{n^2}e^{-E_n/kT_e}+
  \sum_{n^\prime>n} \frac{{n^\prime}^2}{n^2}e^{\Delta E/kT_e}
    A_{n^\prime n} b_{n^\prime}\nonumber\\
    +& \sum_{n^\prime} \frac{{n^\prime}^2}{n^2}e^{\Delta E/kT_e}
    j_\nu B_{n^\prime n} b_{n^\prime} + 
  \sum_{n_c}\sum_{n^\prime} q_{nn^\prime} b_{n^\prime} \nonumber\\
  = & b_n \left[\sum_{n^\prime<n} A_{nn^\prime} + \sum_{n^\prime}  j_\nu B_{n n^\prime} +  
    \Gamma_n + \sum_{i}\sum_{n^\prime} n_i^{\text{col}}q_{nn^\prime} +
    \sum_{i} n_i^{\text{col}}S_n^{\text{ion}} \right ]\,,
  \label{eq:CR_bn}
\end{align}
\noindent where $\Delta E= E_{n^\prime}- E_n$ and we have used detailed balance
in the collisional coefficients:
$e^{E_{n^\prime}/kT}g_{n^\prime}q_{n^\prime n} = e^{E_n/kT}g_nq_{nn^\prime}$.

It is clear from equations (\ref{eq:departures}) and (\ref{eq:emiss1}) that
\begin{equation}
  j_{nn^\prime} = b_n j_{nn^\prime}^{\text{LTE}}\,,
  \label{eq:emiss2}
\end{equation}
\noindent where $j_{nn^\prime}^{\text{LTE}}$ is the emissivity when the level
of the population of the upper $n$-shell is in LTE.

\begin{figure}
  \begin{center}
    \includegraphics[width=0.5\textwidth,clip]{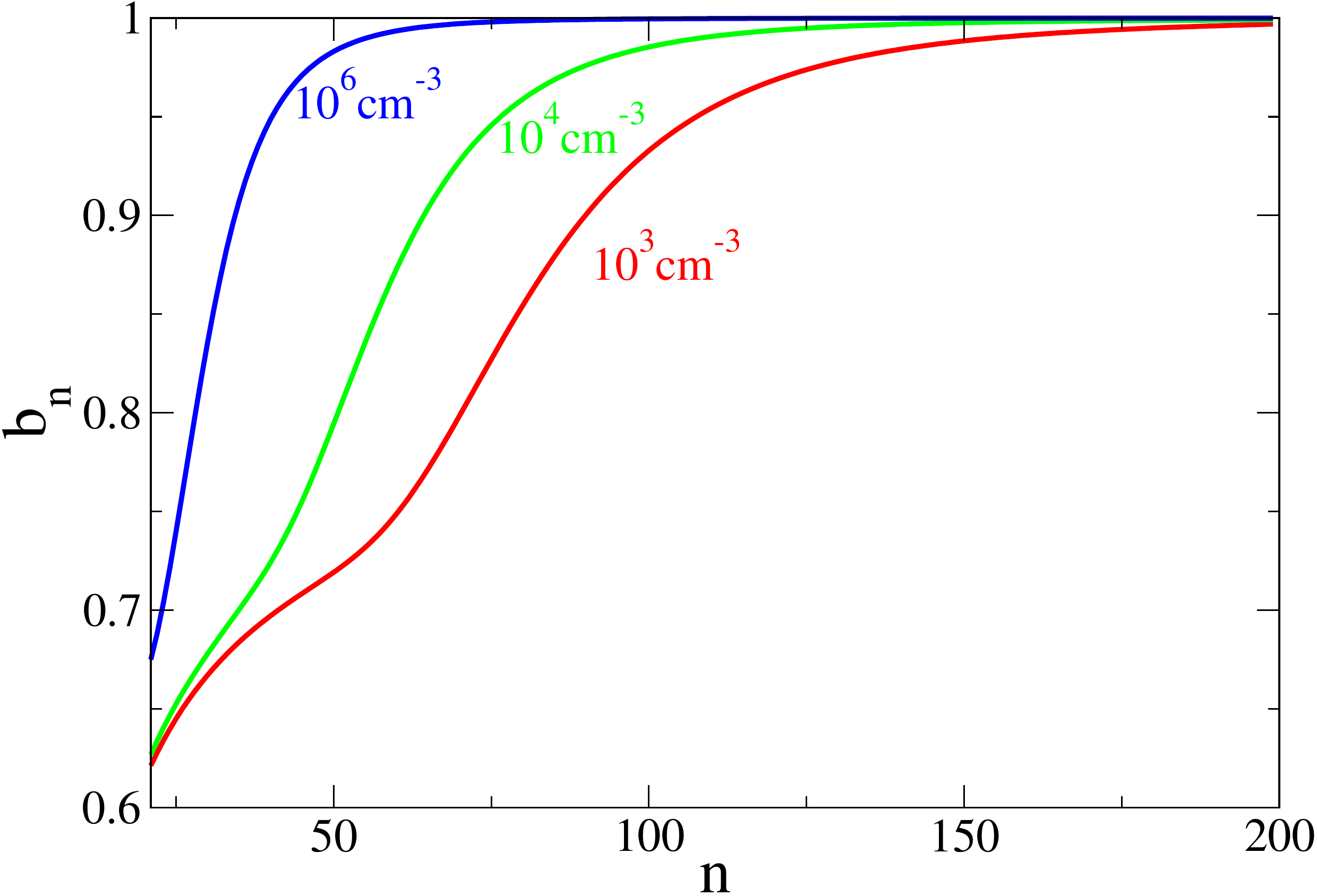}
    \caption{\label{f:bnvsdens} Departure coefficients obtained for a layer of
      gas illuminated by monochromatic radiation (see text for details and
      explanation).}
  \end{center}
\end{figure}

Departure coefficients and populations will depend on both the radiative and
collisional terms in the systems of equation (\ref{eq:CR}) or (\ref{eq:CR_bn}), 
which are functions of the collider temperature and density. In Fig.
\ref{f:bnvsdens}, the $n$-averaged departure coefficients are plotted versus
$n$ ($20<n<200$) for different densities for the plane parallel layer of the
gas model in Section \ref{sec:1layer}. As electron-impact collisional
transitions rates values increase with the principal quantum number as $n^4$, a
lower rate than radiative lifetimes ($n^5$), collisional transitions together
with the three body recombination will drive the high-$n$ levels to LTE. At
higher densities, progressively lower $n$-shells are in LTE. At lower
densities, the contribution of collisional processes is lower, letting the
population radiatively decay to lower levels.

\subsection{One-layer model}
\label{sec:1layer}

To study the different $n$-changing collisional rate formulas described in
Section \ref{sec:datasets}, we have simplified the problem to one illuminated
layer of gas, composed either of hydrogen or of hydrogen and helium, where the 
abundance of the latter is $\frac{n_{\text{He}}}{n_{\text{H}}} = 0.1$ to
resemble cosmological abundances. We have illuminated this gas with
monochromatic radiation of $E=2$Ryd, and forced the gas to a constant electron
temperature and density. We have used a development version of the spectral
code Cloudy \citep{CloudyReview} ({\it nchanging} branch, revision r12537),
where we have implemented the different datasets, to solve the CR equations
(\ref{eq:CR}) and to predict the spectrum. We have used 20 $nl$-resolved
states and unresolved, ``collapsed'', levels up to $n=200$ using equation
(\ref{eq:Nnl}). We have assumed case B \citep{Baker1938} where the Lyman lines
are reabsorbed by the gas which is optically thick for this wavelength. We
have shut down photo-ionization from the $2s$ metastable level and removed the
continuous pumping. We have turned off the escape of photons following the
scattering by a free electron. These operations are done to prevent our
results from being affected by these effects and leading to possible
misinterpretations of our analysis. Photoionization and pumping cause a
re-distribution of the populations of upper levels varying the intensity of
the lines. Escaping photons can reduce the ionization degree. The departure
coefficients in Fig. \ref{f:bnvsdens} are obtained using this model for a
layer of gas composed only of hydrogen. In this case we have used LB98 dataset
for collisional excitation of high Rydberg levels.  

\subsection{Optical depths}
\label{sec:opdepth}

\begin{figure}
  \begin{center}
    \includegraphics[width=0.5\textwidth,clip]{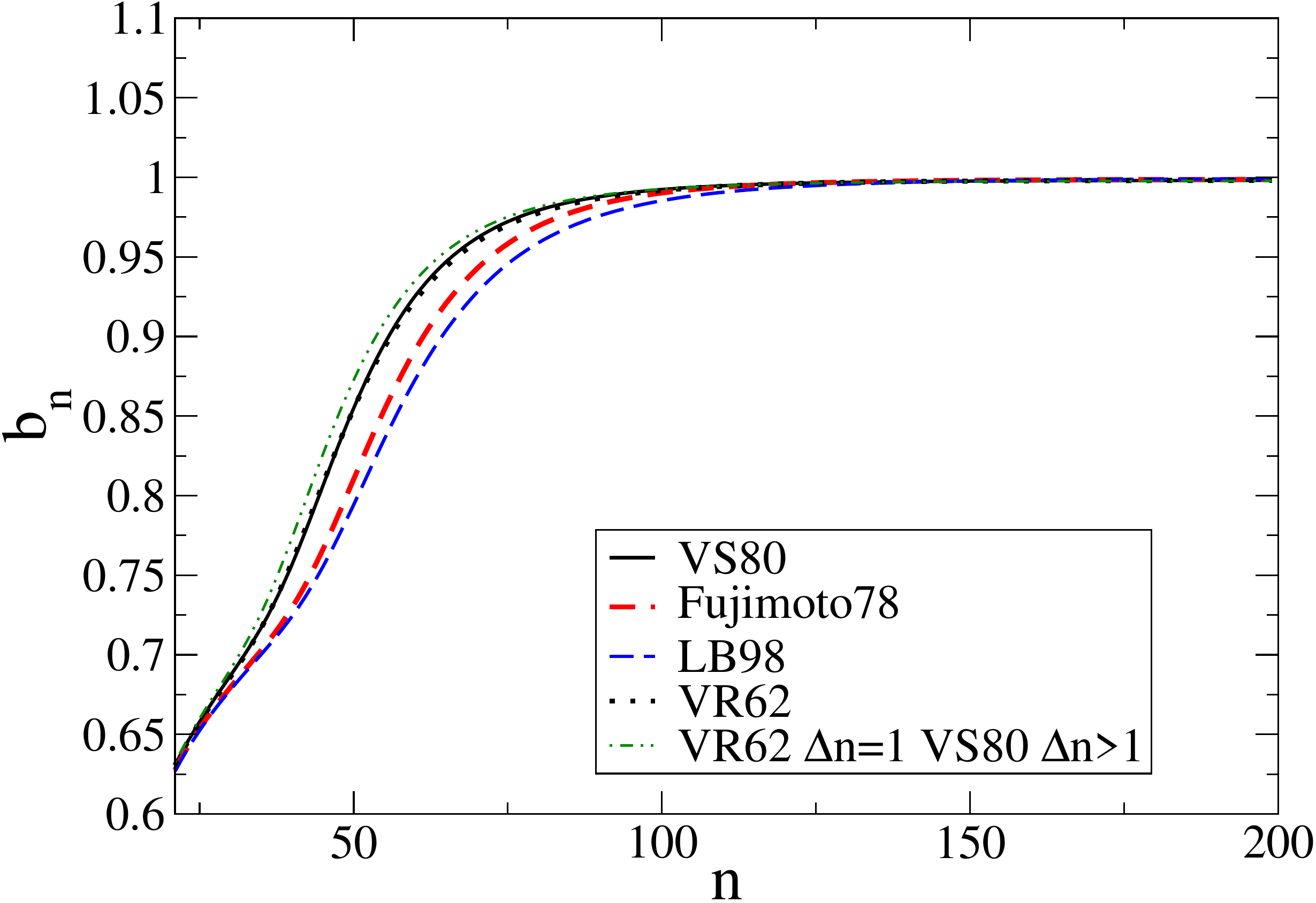}
    \caption{\label{f:bnvsdata} Departure coefficients of H$^0$ obtained for a 
      layer of gas ($T_e=10^4$K and $n_e=10^4\text{cm}^{-3}$) illuminated by 
      monochromatic radiation. The different curves correspond to the
      $n\to n^\prime$ electron collisional datasets of Fig. \ref{f:nnm1},
      except for the double-dotted-dashed green line where the collisional sets
      are taken from VR62 for $\Delta n=1$ and VS80 for $\Delta n>1$ (see text
      for details).}
  \end{center}
\end{figure}

The emitted intensity is obtained solving the equation of radiative transfer
\begin{equation}
  \frac{dI_\nu}{ds} = -\kappa_\nu I_\nu + j_\nu\,,
  \label{eq:RT}
\end{equation}
\noindent where $I_\nu$ is the input intensity, $s$ is the path length through
the gas, and $\kappa_\nu$ is the absorption coefficient. The solution, in
absence of input radiation \citep[see e.g.][]{Rohlfsandwilson2000}, is:
\begin{equation}
  I_\nu = B_\nu\left(1 - e^{-\tau_\nu(s)}\right)\,,
  \label{eq:RTsol}
\end{equation}
\noindent with
\begin{equation}
  \tau_\nu(s)= \int_s^{s_0} \kappa_\nu ds
  \label{eq:optdepth}
\end{equation}
\noindent the optical depth, and $B_\nu=j_\nu/\kappa_\nu$. The absorption
coefficient of a recombination line $n\to n^\prime$ depends on the departure
coefficients of the transition levels and the correction for stimulated
emission \citep{Goldberg1966}:
\begin{equation*}
  \left(1-\frac{b_n}{b_{n^\prime}}e^{-h\nu/kT}\right)\,.
\end{equation*}
$\kappa_\nu$ can be expressed as
\begin{equation}
  \kappa_\nu = \beta_{n^\prime}b_{n^\prime}\kappa^{\text{LTE}}_\nu\,,
  \label{eq:kappa}
\end{equation}
\noindent where $\kappa^{\text{LTE}}_\nu$ and $\beta_n$ can be written as
\citep{Brocklehurstandseaton1972}:
\begin{subequations}
  \begin{align}
    \kappa^{\text{LTE}}_{\nu} = & n_+n_e\frac{hc^2}{16\pi\nu}
    \left(\frac{h^2}{2\pi mkT}\right)^{3/2}
    \frac{g_{n^\prime}}{kT}e^{E_j/kT}A_{nn^\prime} \label{eq:kappaTE}\\
    \beta_n \eqsim & 1-\left(\frac{kT}{\text{I}_{\text{H}}}\right)\left(\frac{n^3}{2}\right)
    \frac{d\ln b_n}{dn} \label{eq:beta}\,,
  \end{align}
\end{subequations}
\noindent with $\nu$ the frequency of the transition $n\to n^\prime$.
$\beta_n$ is the correction of the absorption coefficient for stimulated
emission and depends on both the departure coefficients and their slope with
the principal quantum number.  The departure coefficients, obtained by solving
the CR equations (\ref{eq:CR_bn}) for a pure hydrogen gas using the different
collisional datasets of Section \ref{sec:datasets}, are given in Fig.
\ref{f:bnvsdata} as a function of the principal quantum number. All curves are
similar and slightly reflect the differences shown in Fig. \ref{f:nnm1}.

\begin{figure}
  \begin{center}
    \includegraphics[width=0.5\textwidth,clip]{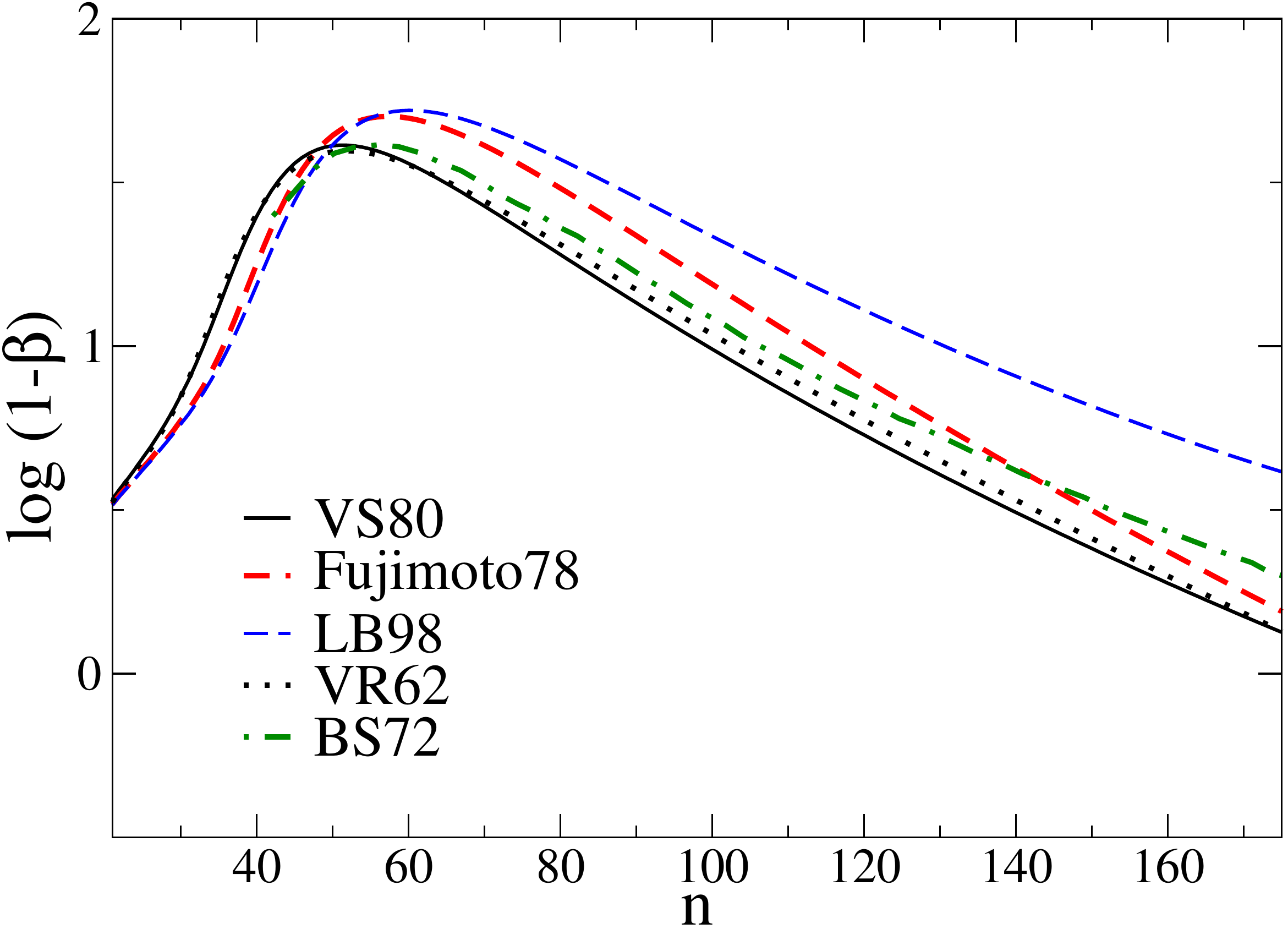}
    \caption{\label{f:betas} $\log(1-\beta)$ coefficients obtained for a layer
      of gas ($T_e=10^4$K and $n_e=10^4\text{cm}^{-3}$ illuminated by
      monochromatic radiation). BS72 are the $\beta_n$ factors published
      in figure 2 of \citet{Brocklehurstandseaton1972}. The other curves
      correspond to the $n\to n^\prime$ electron collisional data of Fig.
      \ref{f:nnm1}. }
  \end{center}
\end{figure}

In Fig. \ref{f:bnvsdata}, larger excitation rate-coefficients produce a
marginally steeper slope in the departure coefficients due to the faster decay
probabilities of the lower levels. VS80 and VR62 give similar departure
coefficients despite their different rates for $\Delta n=1$ and $\Delta n=2$.
The cause of that is related with the fact that the rate coefficients for
$\Delta n>3$ for VR62 are lower than the VS80 ones, compensating for the
differences of lower $\Delta n$ transitions. We have checked this by
performing a hybrid calculation where the rate coefficients for $\Delta n>1$
are taken from VS80 fits while rates coefficients with $\Delta n=1$ are from
VR62 formula. The results are plotted in the double-dotted-dashed green line
of Fig \ref{f:bnvsdata}, which is steeper than the pure VR62 and VS80. The
$\beta_n$ factors depend on the slope of the departure coefficients through
equation (\ref{eq:beta}). The $\beta_n$ for this single-layer model are given
in Fig. \ref{f:betas}. Population inversion, which is a condition for
stimulated radiation, happens for $\beta_n<0$ ($\log(1-\beta_n)>0$). The
maximum of $\log(1-\beta_n)$ corresponds to the point where the population
inversion is highest. This point shifts for the different datasets in
correlation with the value of the collisional rate coefficients, because of
the collisional distribution of the populations. As manifested in Fig.
\ref{f:betas}, $\beta_n$ factors turn out to be very sensitive to the
differences in collisional data, which will affect the absorption coefficients,
and so the optical depth and the observed spectrum. LTE is reached at
$\beta_n = 1, b_n =1$ ($\log(1-\beta_n)=-\infty$). All datasets tend to LTE
when $n$ is high. However, big differences in the value of $\beta_n$ arise
from the different slopes of the departure coefficients of Fig.
\ref{f:bnvsdata}, due to the different collisional (de-)population
probabilities. LTE will finally be ensured by the cooperation of collisional 
(de-)excitation and collisional ionization/three body recombination processes. 
It is important that our model atom is big enough to include the collisional 
effects of the continuum coupled to the highest levels. Their population will 
be transferred to lower $n$ by collisional de-excitation, which is dominant. A
brief discussion about the importance of the completeness of the model atom is
included in appendix \ref{sec:atommodel}.

\citet{Brocklehurstandseaton1972} studied the functional dependence of the
$\beta_n$ factors with temperature using the departure coefficients from 
\citet{Brocklehurst1970}. Their model is a theoretical calculation of an
homogeneous pure hydrogen nebula that is very similar to our model. 
\citet{Brocklehurst1970} used the impact parameter theory of
\citet{Seaton1962} for the dominant $\Delta n=1$ collisional transitions, and
his values are close to Fujimoto78, who fitted these values (see Section
\ref{sec:F78}). Their $\beta_n$ are the dot-dashed curve represented in Fig.
\ref{f:betas}. We could reproduce their results using the excitation rate
coefficients formula from \cite{Percival1978}. This formula give similar rate
coefficients to the \citet{Brocklehurst.Nat.1970} datasets used originally by
\citet{Brocklehurstandseaton1972}\footnote{Later
  \citet{Brocklehurstandsalem1977} used \citet{Gee1976} formulas in their code
  for the calculation of departure coefficients.}.

\section{Impact on Optical Recombination lines}
\label{sec:optlines}

\begin{figure}
  \begin{center}
    \includegraphics[width=0.5\textwidth,clip]{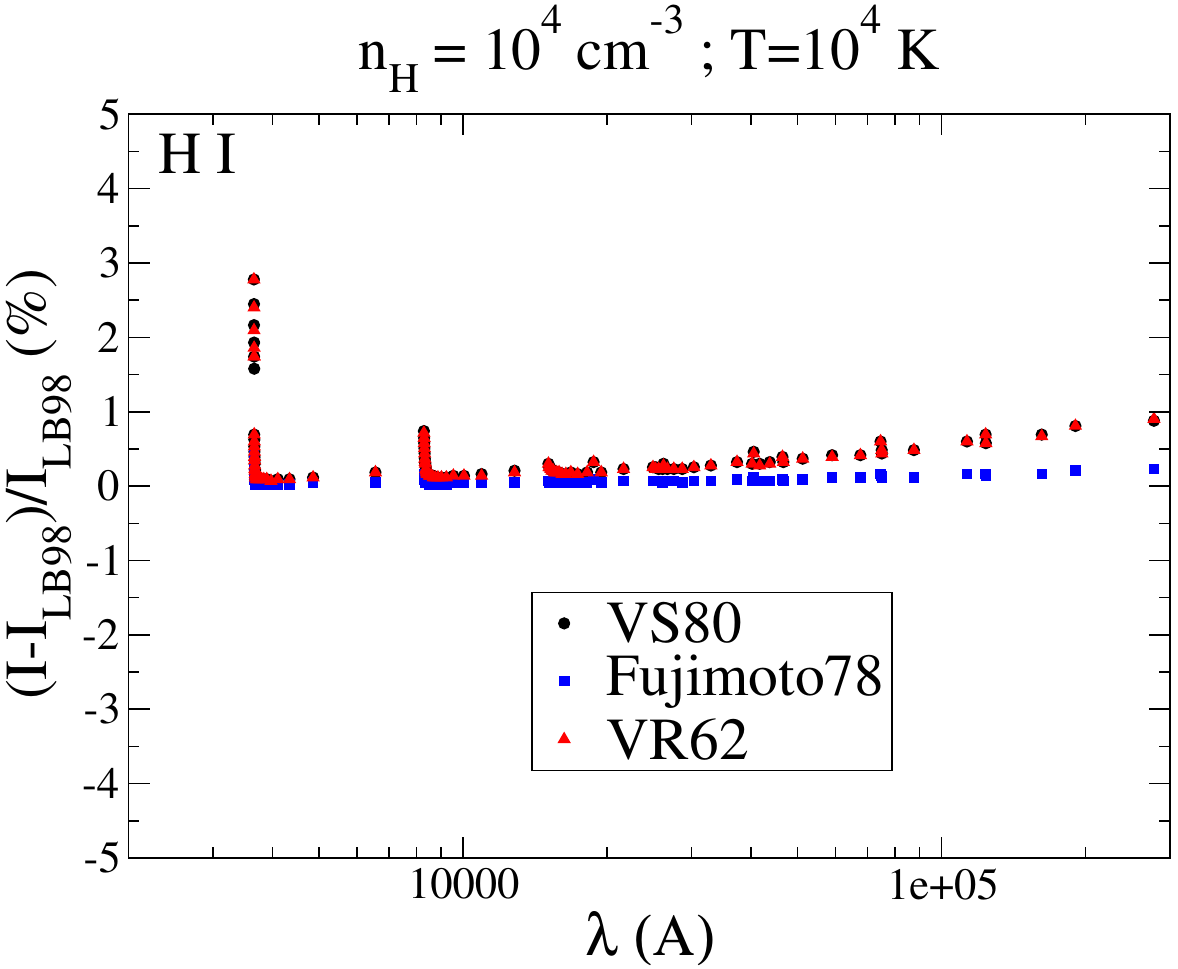}
    \caption{\label{f:optratiosh} Ratios of H I emitted intensities for
      different datasets for a pure hydrogen gas
      ($n_{\text{H}}=10^4\text{cm}^{-3}$, $T_e=10^4$K) illuminated by an
      ionizing monochromatic radiation of 2Ryd. The intensities are normalized
      to the LB98 results and given in percent units.}
  \end{center}
\end{figure}

\begin{figure}
  \begin{center}
    \includegraphics[width=0.5\textwidth,clip]{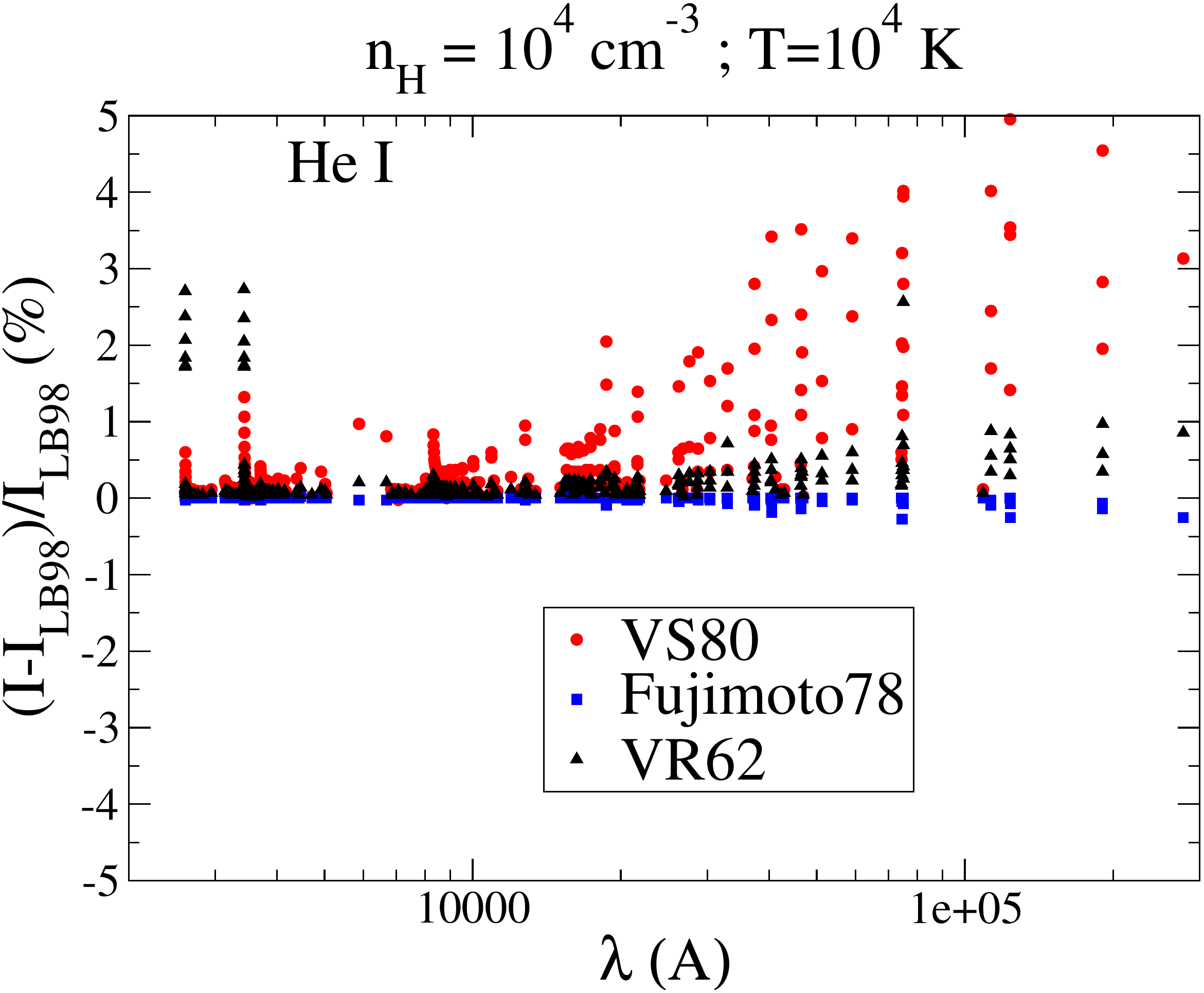}
    \caption{\label{f:optratioshe} Ratios of He I emitted intensities for a
      hydrogen and helium gas (abundance
      $\frac{n_{\text{He}}}{n_{\text{H}}}= 0.1$,
      $n_{\text{H}}=10^4\text{cm}^{-3}$ and $T_e=10^4$K) illuminated by an
      ionizing monochromatic radiation of 2Ryd, enough to ionize both H and He.
      The intensities are normalized to the LB98 results and given in percent
      units.}
  \end{center}
\end{figure}

High-$n$ excitation processes can affect optical recombination lines due to
the cascade of the electrons to lower levels. In Fig. \ref{f:optratiosh} we
show the percentage difference of intensities of H~I recombination lines, with
principal quantum number $n=1$ to $n=20$, for the datasets discussed in
Section \ref{sec:datasets} with respect to LB98. We have used the one layer
model discussed in Section \ref{sec:1layer}, with pure hydrogen gas
($n_\text{H}= 10^4 \text{cm}^{-3} \simeq n_e$), and electron temperature
($T_e = 10^4$K). Fig. \ref{f:optratioshe} shows the He~I recombination lines
for a gas composed of hydrogen and helium as in Section \ref{sec:1layer}. At
these densities the effect of the cascading electrons is highest, as
excitation and de-excitation collisions are competing with radiative processes
(see Paper I). At lower densities radiative processes are dominant and at
higher densities ($n_{\text{H}}\sim 10^{10}\text{cm}^{-3}$) collisional
processes will bring the higher Rydberg levels to LTE, and no differences will
be evident on the optical recombination lines. Excitation rate coefficients up
to $n=5$ have been taken from the R-matrix calculations of
\citet{Anderson2000} for hydrogen and from \citet{Bray2000} for helium in all
the simulations.

In Fig. \ref{f:optratiosh} the different series of lines $\Delta n= n_u - m $ 
(for a fixed principal quantum number of the lower level, $m>1$) can be seen.
Inside  each of the series, the lower wavelengths correspond to higher energy
leaps where $n_u$ is higher, and higher wavelengths correspond to lower
$\Delta n$.  The deviation is bigger at the short-wavelength end of the series,
because the lines come from levels that have higher $\Delta n$. Deviations
also exist at the long-wavelength end of each series, due to the electron
cascades from higher levels. This is seen in the Balmer series ($m=2$, the
leftmost series of values) in Fig. \ref{f:optratiosh}, where the greatest
disagreement is at lower wavelengths, i.e., higher $\Delta n$. Deviations for
$\Delta n\to1$, are more significant for higher series ($m>2$), and grow with
$m$. Note that even when the higher disagreements are produced for $n>5$ (the
limit of R-matrix calculations), the lines from transitions with upper levels
lower than $n_u=5$ are also slightly affected. In Fig. \ref{f:optratiosh}, the
H$_\alpha$ and H$_\beta$ lines are barely affected and their intensities are 
$\sim$0.2\% higher than the LB98 results. In Figure \ref{f:optratioshe}, the 
situation is less clear as the lines from the singlet and triplet spin systems 
mix, but the different series can be readily distinguished, and the trend is
similar to the H~I case with He~I ($n=4\to2$) disagreeing by no more than 1\%.

From Figs \ref{f:optratiosh} and \ref{f:optratioshe}, the effect of higher
excitation rates in the higher Rydberg levels is to re-distribute the 
populations, thus increasing the impact relative to LB98 datasets in the
intensity of radiative cascades. Some astronomical determinations require high
precision. In the case of the determination of the primordial helium abundance
obtained from He~I and H~I recombination line ratios
\citep[see for example][]{Izotov1997}, a precision better than 1\% is needed
\citep{Olive2000,Izotov2014}, the cascade effect could be influencing only if
the lines corresponding to higher $n$ transitions of a series are used, or for
wavelenths longward of $\sim$30000\AA ~in the He I recombination spectrum.

We expect from these results that, in general, optical recombination
observations will not be affected by the reported uncertainties on high
Rydberg collisions. The non ideality of the conditions of astrophysical plasmas
(local velocity and density fluctuations) will smear even more the effect of
the cascade from high Rydbergs making these differences practically
unobservable.

\section{Effect On Observations of Astronomical Nebulae}
\label{sec:models}

In order to test how the collisional uncertainties propagate to predictions, we
have modeled the spectrum of two different astronomical systems. In the first
case, in Section \ref{sec:agn}, we simulate a typical BLR illuminated by an AGN
with a power-law Spectral Energy Distribution (SED). In the second case,
described in Section \ref{sec:orion}, we explore the effects these different 
collisional data have on a H~II nebula, such as the Orion Blister. The 
calculations have been done with the development version of Cloudy 
\citep{CloudyReview} ({\it nchanging} branch, revision r12537).

\subsection{BLR model}
\label{sec:agn}

Radio recombination lines can be used to probe AGNs' continuum source due to
their minimal dust attenuation \citep{Scoville2013} and to differenciate them
from Startburst galaxies by using the ratio of He II to H I intensities from
RRL. The RRL from the BLR could be detectable using the capabilities of the
SKA radio-telescope \citep{Manti2016,Manti2017}. In this Section, we study how
the different datasets affect our simulations of a cloud irradiated by an
AGN-type spectral energy distribution. In order to do this, we will first
study how the chosen SED and density affect the equilibrium balance, and thus
the emission of the RRL. Then, we will analyze the effect of the different
datasets on the opacity and intensity of RRLs.

\subsubsection{Dependence of the results on the SED}
 
\begin{figure}
  \begin{center}
    \includegraphics[width=0.5\textwidth,clip]{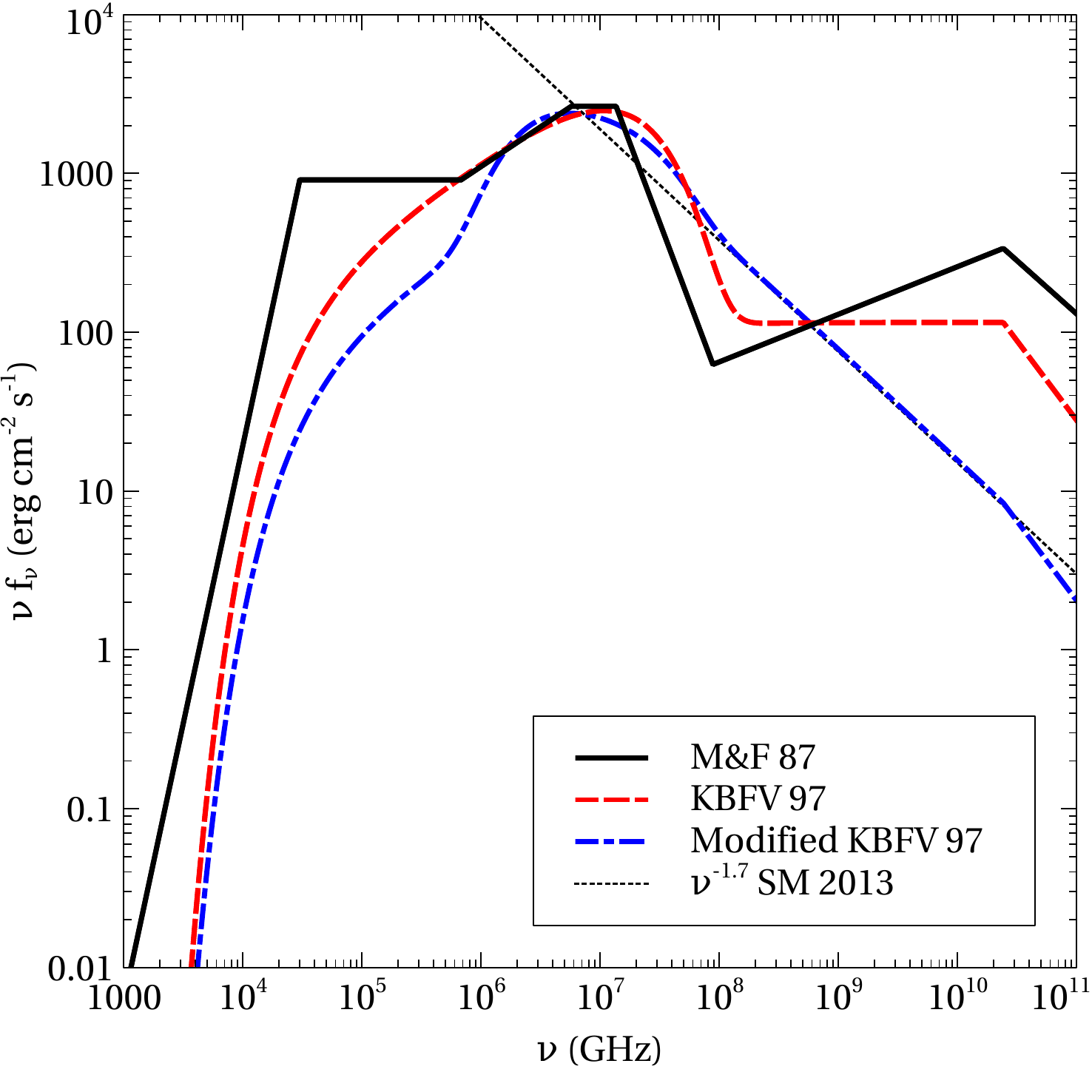}
    \caption{\label{f:agnsed} Incident flux ($\nu f_\nu$) used in our BLR
      models. The legends refer to the following SEDs: M\&F 87:
      \citet{Mathews1987}; KBFV 97: \citet{KoristaBaldwin1997};
      SM 2013: \citet{Scoville2013}. The assumed ionization parameter is
      $\log{U}=-1.5$. For reference, the ionization threshold for hydrogen is
      $\nu=3.5\times 10^7$~GHz.}
  \end{center}
\end{figure}

\citet{Scoville2013} simulate the SED of an AGN source by using a power law in
agreement with \citet{AGN3}:
\begin{equation}
f_\nu \propto \nu^{-\alpha}\,,
\label{eq:powerlaw}
\end{equation}
\noindent where $\alpha=1.7$. For our study, we have used three different SEDs,
shown in Figure \ref{f:agnsed}, in order to check how the incident ionizing
flux in the UV affects the H$n\alpha$ emission and optical depth. The general
shape of AGNs SED is described by the SED proposed by
\citet[hereafter M\&F 87]{Mathews1987}, which describes a ``bump'' of the
incident radiation in the UV, and a cut-off at higher frequencies. A more
complicated formula was proposed by
\citet[hereafter KBFV 97]{KoristaBaldwin1997}:
\begin{equation}
\label{eq:agncon}
f_\nu   = \nu ^{\alpha _{uv} } \exp \left( { - h\nu /kT_{BB} } \right)\exp
\left( { - kT_{IR} /h\nu } \right)\; + a\nu ^{\alpha _x }\,. 
\end{equation}
The Big Bump component in the UV is assumed to have an infrared exponential
cutoff at $kT_{IR} = 0.01 \mathrm{Ryd}$. The coefficient $a$ is adjusted to
produce the correct X-ray to UV ratio. The X-ray power law is only added for
energies greater than 1.36 eV \footnote{Here we follow the custom of using eV
  units for the X-ray spectrum. 1.36 eV correspond to $3.29\times10^5$GHz and
  100keV= $2.42\times10^{10}$GHz.} to prevent it from extending into the
infrared, where a power law of this slope would produce \emph{very} strong
free-free heating; and less than 100~keV. Above 100 keV the continuum is
assumed to fall off as $v^{-2}$. For KBFV 97 $T_{BB}= 1.5\times10^5K$,
$\alpha_{uv}=-0.5$, and $\alpha_x=-1$.

M\&F's and KBFV's SEDs differ from the exponential law model used by
\citet{Scoville2013} in the high energy region. The shape of the high energy
cut-off of the different SED could affect the ionization fraction of hydrogen
and the length of the ionization phase, so we expect slightly different RRL
emission. We have created a different SED to reproduce the $\nu^{-1.7}$ power
law, while conserving the drop of the flux  at low frequencies in order to
avoid an unrealistic infrared radiation field, using the indices
$T_{BB}= 10^6K$, $\alpha_{uv}=-0.5$ and $\alpha_x=-1.7$ in equation
(\ref{eq:agncon}) , shown in Fig. \ref{f:agnsed} as a blue dot-dashed line.

Simulations were carried out for H atoms with levels up to $n=205$ of which 
$n=1-5$ are $l$-resolved (200 unresolved levels). At the lowest densities
considered here, the $n$-shells immediately higher than $n=5$ will not be
statistically populated in the $l$-subshells, but we are focused in emission
of lines at higher $n\sim30$ and we do not expect this to affect our results.
We have stopped our calculation when the ratio of the electron to hydrogen
densities is 0.5, meaning that half of the hydrogen is
recombined\footnote{Cloudy command \emph{stop eden 0.5}}. This ensures that we
are obtaining the emission of the ionized phase of the cloud irradiated by the
AGN. If a neutral phase exists beyond the ionization front, it might affect the
observed intensities and optical depths and depend strongly on the SEDs. This
will be the scope of future work.

\begin{figure}
  \begin{center}
    \includegraphics[width=0.5\textwidth,clip]{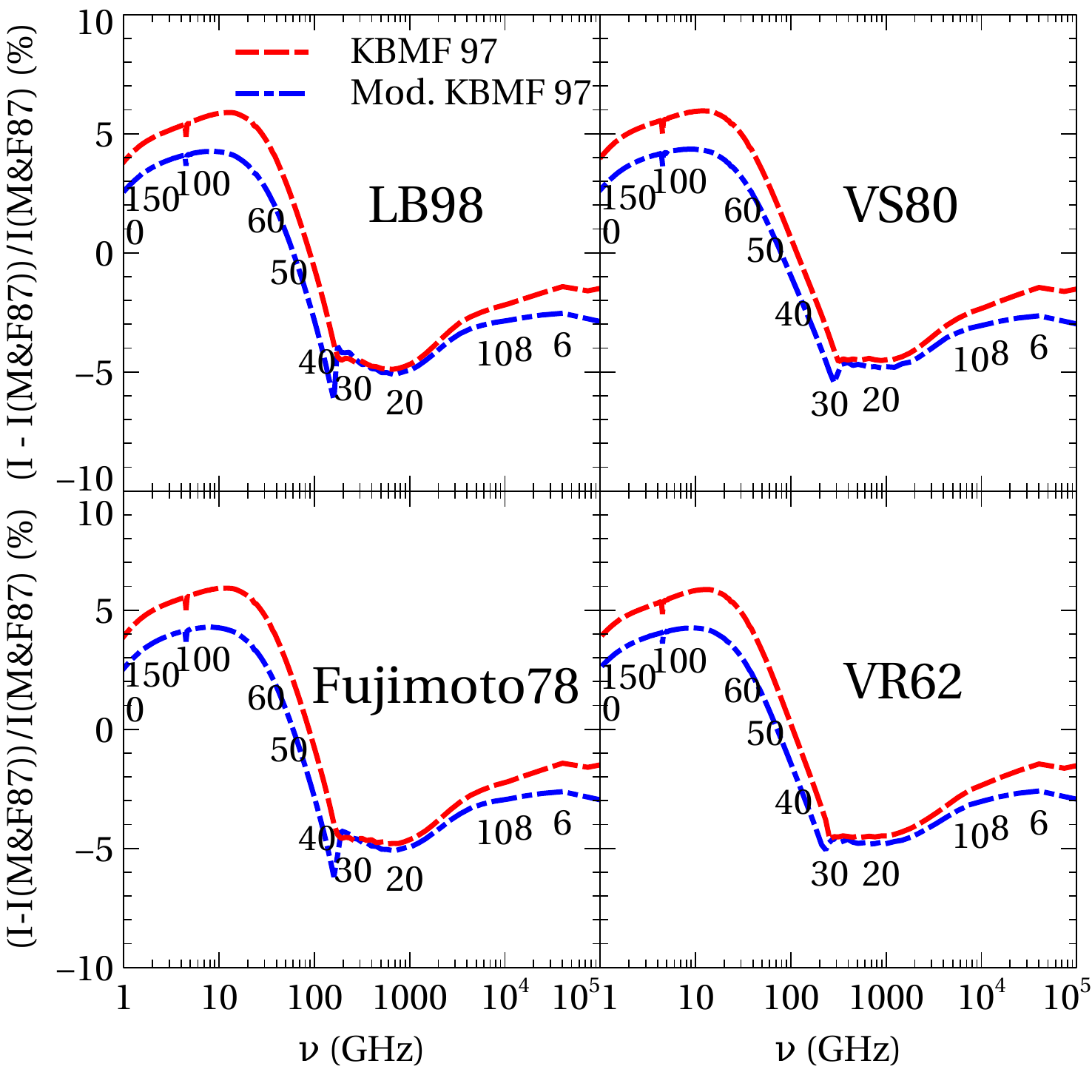}
    \caption{\label{f:sedemiss} Percentage difference of H$n\alpha$ RRL 
      emission with respect to the calculations using M\&F 87 SED for the
      KBFV 97 and the modified KBFV 97 SED shown in Fig. \ref{f:agnsed}.
      Calculations where done for density $n=10^8\text{cm}^{-3}$. Labels
      indicate the level $n$ of the H$n\alpha$ lines. Each panel correspond to
      each of the different collisional datasets of Fig. \ref{f:nnm1}.}
  \end{center}
\end{figure}

Relative RRL emission with respect to M\&F 87 SED are shown in Fig.
\ref{f:sedemiss} for the different SEDs considered here, at densities 
$n_\text{H} =10^8\text{cm}^{-3}$ and for the different datasets considered
here. There is a disagreement of the emission of H$n\alpha$ lines with respect
to the M\&F 87 SED, with a peak localized at $\nu\sim10$ GHz or H$80\alpha$
for all collisional datasets. This is the same for all datasets. The peak is
explained by the amount of the UV photons available that increase the
thickness of the column of ionized gas.

In summary, the shape of the assumed SED can translate to differences in the
emitted intensities like the ones reported in Fig. \ref{f:sedemiss}, when the
UV/X-ray intensity of the SED affects the thickness of the Str\"omgrem sphere.

\subsubsection{Dependence of the results on the BLR density}

\begin{figure}
  \begin{center}
    \includegraphics[width=0.5\textwidth,clip]{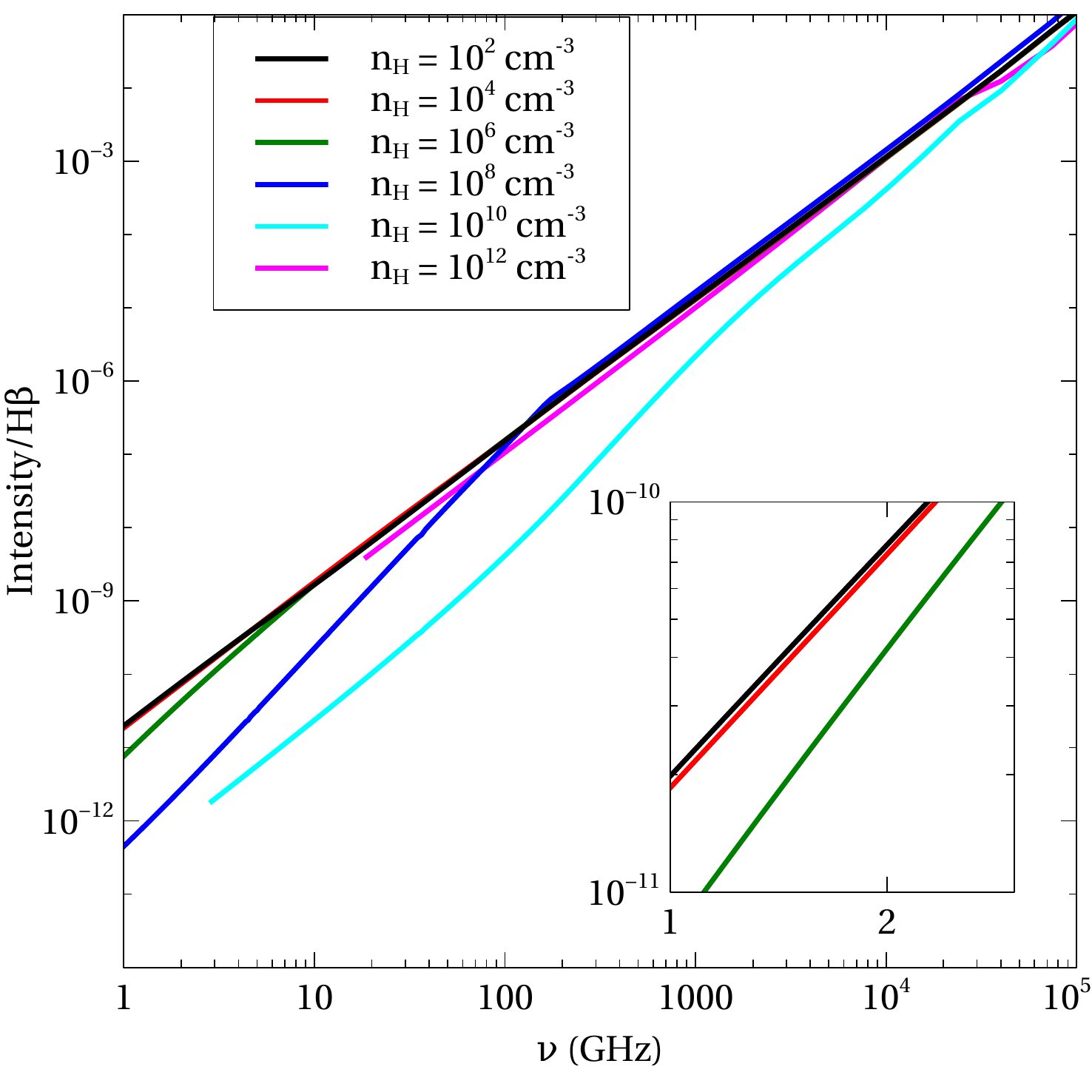}
    \caption{\label{f:agnintensities} H$n\alpha$ intensities normalized to
      H$\beta$ for a layer of gas irradiated by the M\&F 87 SED of Fig.
      \ref{f:agnsed}. Collisional excitation data from LB98 dataset. At
      $n_\text{H} = 10^{10} \text{cm}^{-3}$ and
      $n_\text{H} = 10^{12} \text{cm}^{-3}$ the curves are truncated due to
      continuum lowering. In the inset there is a enlargement of the low
      frequencies region.}
  \end{center}
\end{figure}

\begin{figure}
  \begin{center}
    \includegraphics[width=0.5\textwidth,clip]{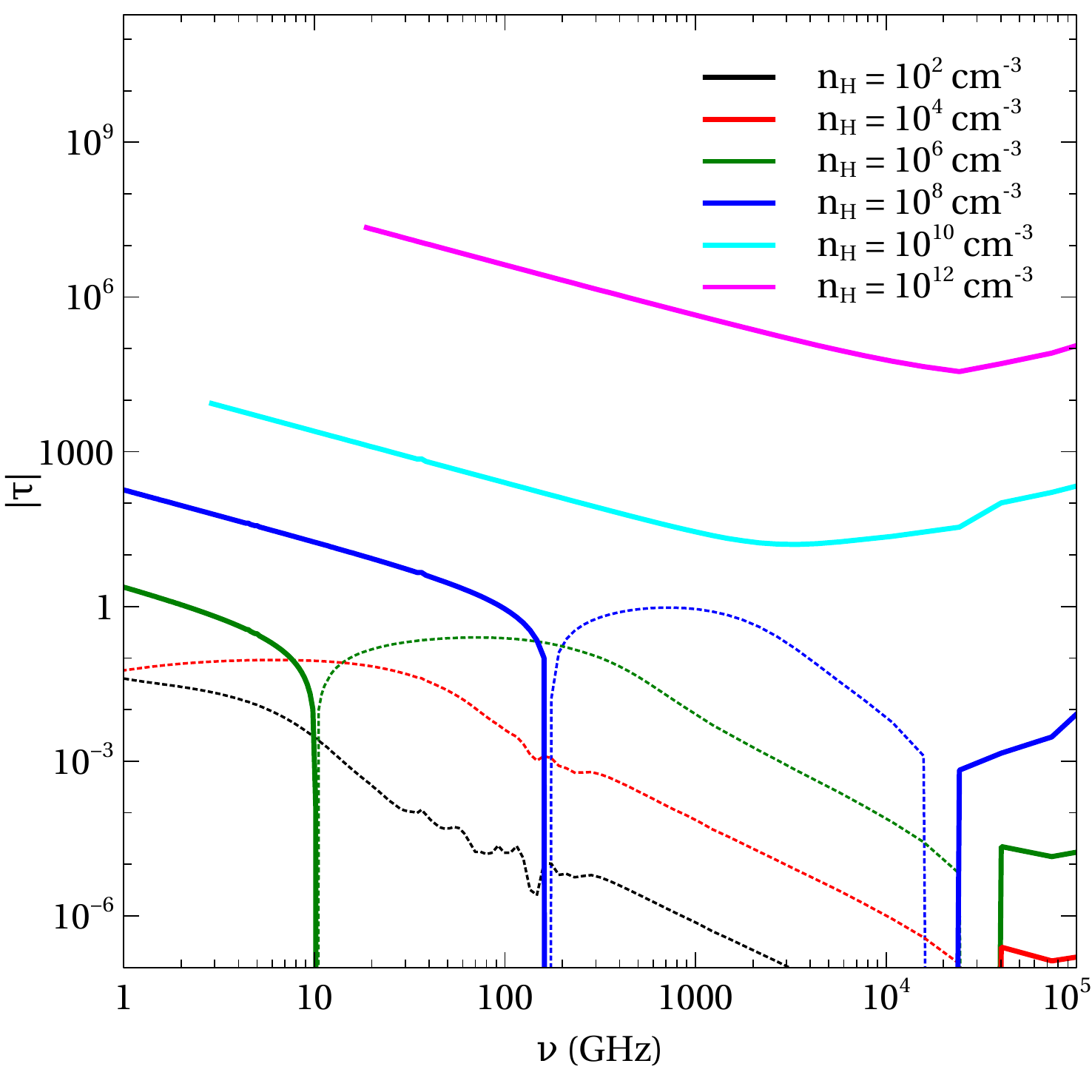}
    \caption{\label{f:agnoptdepth} Absolute values of the optical depths at
      H$n\alpha$ wavelengths for a layer of gas irradiated by the M\&F 87 SED
      of Fig. \ref{f:agnsed}. Collisional excitation data from LB98 dataset.
      Dotted lines correspond to the negative values of $\tau$. At
      $n_\text{H} = 10^{10} \text{cm}^{-3}$ and
      $n_\text{H} = 10^{12} \text{cm}^{-3}$ the curves are truncated due to the
      continuum lowering.}
  \end{center}
\end{figure}

To analyze how the properties of the BLR vary with the gas density
$n_\text{H}$, we have used the SED from M\&F 87 and LB98 $n$-changing
collisional excitation rates, and varied the density of the irradiated gas
while keeping the ionization parameter $U$ \citep{AGN3} constant. The
H$n\alpha$ RRL emitted intensity (normalized to H$\beta$) is given in Fig.
\ref{f:agnintensities} for hydrogen densities running from
$n_\text{H}=10^2 \text{cm}^{-3}$ to $n_\text{H}=10^{12} \text{cm}^{-3}$. For
the lowest densities, the curves $n_\text{H}=10^2 \text{cm}^{-3}$,
$n_\text{H}=10^6 \text{cm}^{-3}$ and $n_\text{H}=10^8 \text{cm}^{-3}$ switch
their position progressively at $\sim300$ GHz, $\sim20$GHz and $\sim2$ GHz. The
origin of this is explained by looking at the optical depths given in Fig.
\ref{f:agnoptdepth}. For frequencies longward of $\nu\sim200$GHz up to
$\sim2\times10^4$ GHz, the H$n\alpha$ transitions are mased for densities
$n_\text{H}=10^2-10^8 \text{cm}^{-3}$ which are all in the same order of
magnitude (note the 13 orders of magnitude range of Fig. \ref{f:agnintensities})
but their intensity ratio with respect to the H$\beta$ line is bigger according
to how strong they are masing. At $\sim300$ GHz, the ionized plasma at
$n_\text{H}=10^8 \text{cm}^{-3}$ becomes optically thick and keeps increasing
its optical depth up to $\tau=1$ at $\sim200$ GHz coinciding with the
inflexion of the intensity curve, which is not amplified anymore by masers, but
rather damped by absorption. The same process happens for
$n_\text{H}=10^6 \text{cm}^{-3}$ at $\sim20$ GHz and again at $\sim1$GHz for
$n_\text{H}=10^4 \text{cm}^{-3}$ (inset of Fig. \ref{f:agnintensities}).
At $n_\text{H}\geq10^{10} \text{cm}^{-3}$ there is no masing in this frequency
range, so the H$n\alpha$ lines are not amplified. This is because, at the
higher electron densities, collisions bring the levels close to LTE and undo
the inversion of the population that causes masing. Note that, for densities
$n_\text{H}=10^{12} \text{cm}^{-3}$, the higher absorption is compensated by the
higher number of emitting atoms producing a flux higher than the one for
$n_\text{H}=10^{10} \text{cm}^{-3}$.

The ranges of negative $\tau$ do not agree with the ones of
\citet{Scoville2013}. These authors, use the absorption coefficients provided by
\citet{Storey1995}, for a range of temperatures and densities. Our calculations
using Cloudy are self-consistent and the temperature and ionization structure
are deduced from the radiation field and the density after solving the CR
equations (\ref{eq:CR}). The cloud is divided in zones and the radiation output
from one zone is the input of the next one \citep[see ][]{CloudyReview}. The
SED distribution in the infrared will contribute to ionize the Rydbergs 
carrying their population far from LTE.

\begin{figure}
  \begin{center}
    \includegraphics[width=0.5\textwidth,clip]{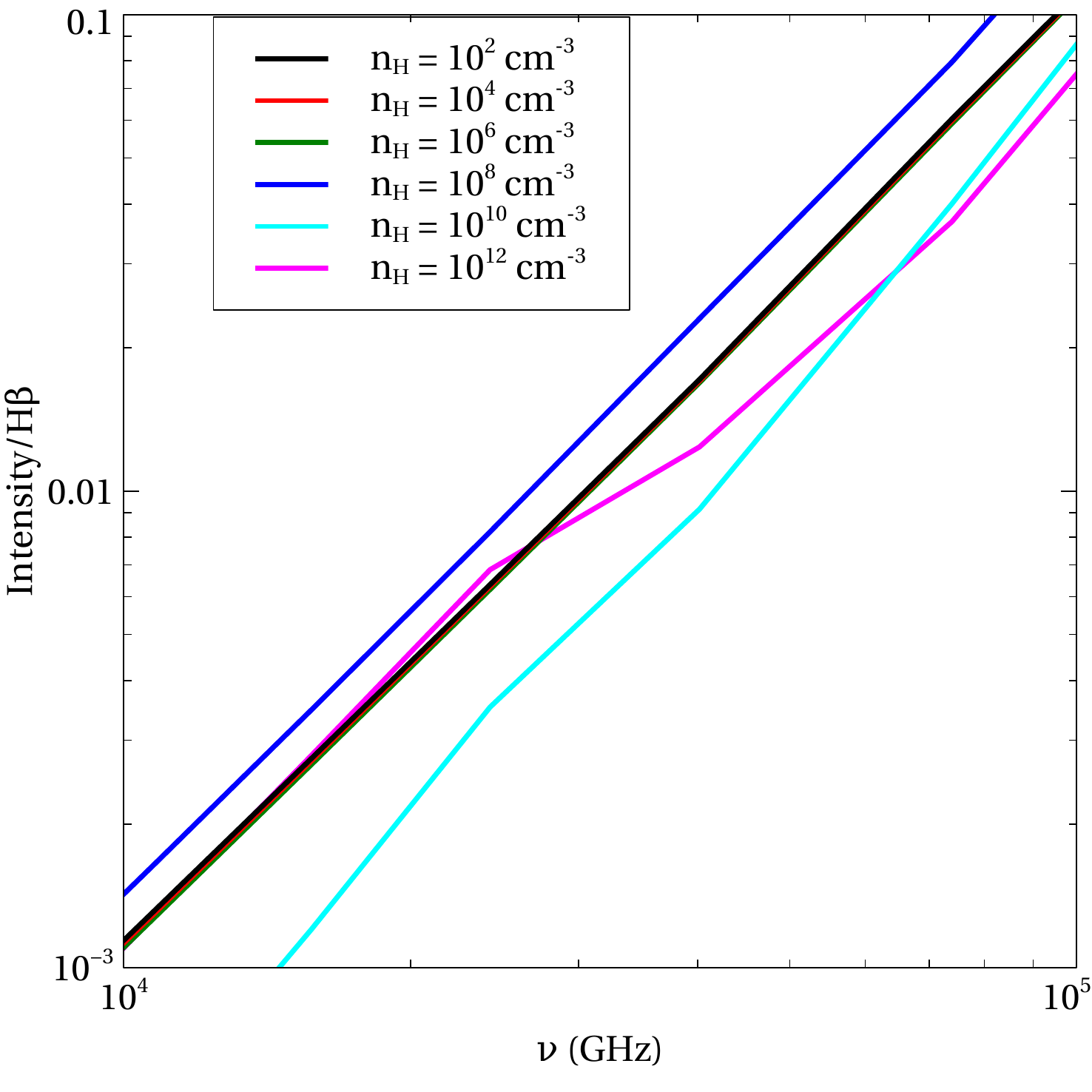}
    \caption{\label{f:inthighnus} Zoom at high frequencies of
      Fig. \ref{f:agnintensities}. }
  \end{center}
\end{figure}

At high frequencies $\nu\sim3\times10^4$ GHz the low density curves in Fig.
\ref{f:agnintensities} are parallel and there is no change even when the
H$n\alpha$ lines are mased (Fig. \ref{f:inthighnus}). The reason for this is
that the optical depths are too low and the plasma is optically thin.
Therefore, the variation of the optical depth is not enough to produce any
changes in the H$n\alpha$ intensities in this range of frequencies where the
number of emitting atoms is determinant.

\emph{Continuum Lowering}. For $n_\text{H} = 10^{10} \text{cm}^{-3}$ and
$n_\text{H} = 10^{12} \text{cm}^{-3}$ the continuum pressure lowers the atom to
the principal quantum numbers $n=132$ and $n=71$ respectively, producing a
discontinuity in the emission. Continuum lowering is treated in Cloudy taking
the minimum $n$ given by continuum lowering by particle packing, Debye
shielding and Stark broadening of the atomic levels using the formulas
suggested by \citet{Bautista00}.

\subsubsection{Dependence of the results on the collisional-excitation 
rate-coefficients}

\begin{figure*}
  \begin{center}
    \includegraphics[width=\textwidth,clip]{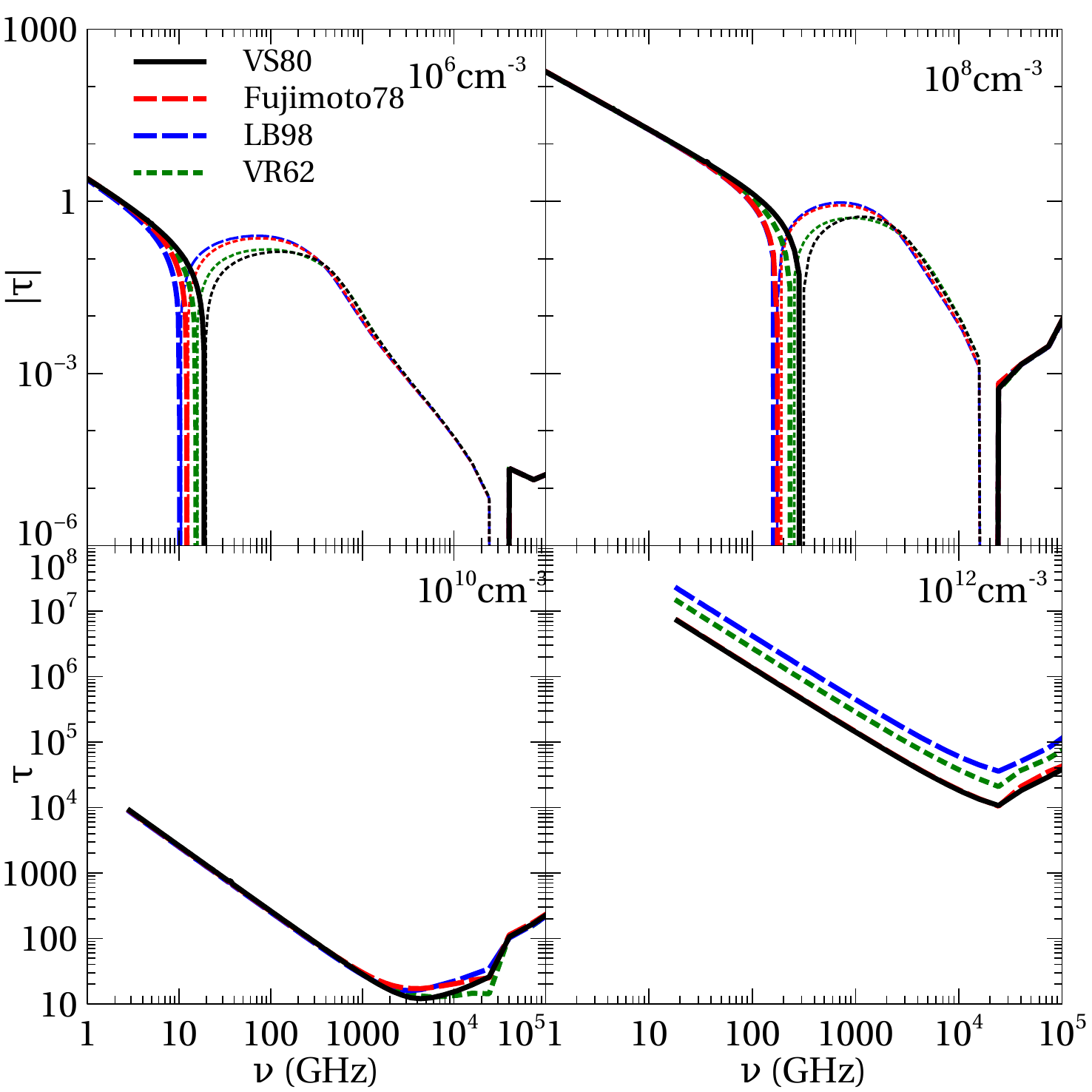}
    \caption{\label{f:agnODcompMF} Comparison of the optical depths obtained
      with the different collisional datasets of Fig. \ref{f:nnm1} at
      H$n\alpha$ wavelengths for a layer of gas irradiated by the M\&F 87 SED
      of Fig. \ref{f:agnsed} at different densities. Thick lines correspond to
      $\tau>0$ while thin lines to $\tau<0$. Labels correspond to the datasets
      from Fig. \ref{f:nnm1}.}
  \end{center}
\end{figure*}

Increasing the value of collisional excitation rates for high $n$ is partially
analogous to increasing the density, because the number of collisions is
compensated with a greater probability of excitation per collision. However,
higher density also increases the opacity to the incident radiation and can
produce greater diffuse emission field. This effect cannot be reproduced by
changing the values of collisional excitation rates. Nevertheless, the
equilibrium produced by redistribution of population in the high Rydberg
levels by a more efficient collisional excitation shifts the masing range to
lower levels, having an important effect on the emission of RRL. This is
indeed what can be seen in Fig. \ref{f:agnODcompMF}, for the different SED
considered above. The masing range (thin dashed lines at
$n_\text{H}=10^6 - 10^8 \text{cm}^{-3}$) gets down to lower frequencies (higher
$n$ of the H$n\alpha$ lines) for the datasets with lower values of excitation
rate coefficients, as their population are far from LTE yet. The electron
temperature of the fully ionized gas, from where most of the radio
recombination emission comes from, is $T_e\sim15000$K. At higher densities
($n_\text{H}=10^{10} - 10^{12} \text{cm}^{-3}$) the optical depths are so high
that most of the levels of the system can be considered to be close to LTE and
all the datasets result in parallel curves depending on the differences in
collisional rates. Similar results with small variations are obtained for the
other SED considered above.

\begin{figure*}
  \begin{center}
    \includegraphics[width=\textwidth,clip]{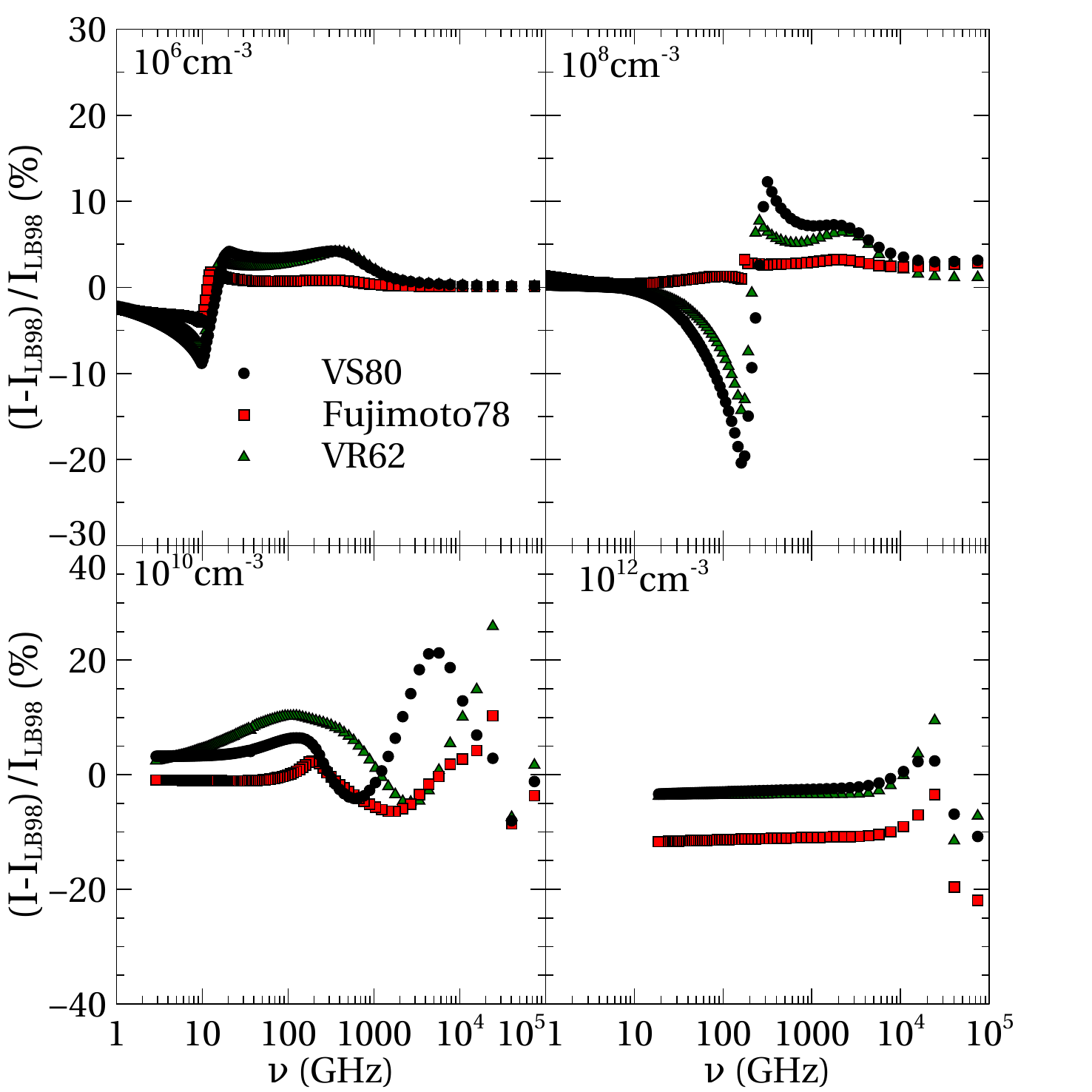}
    \caption{\label{f:intcompMF} Comparison of the intensities (with respect to
      LB98) obtained with the different collisional datasets of Fig.
      \ref{f:nnm1} at H$n\alpha$ wavelengths for a layer of gas irradiated by
      the M\&F 87 SED of Fig. \ref{f:agnsed}. Labels correspond to the datasets
      from Fig. \ref{f:nnm1}}
  \end{center}
\end{figure*}

The differences in optical depths propagate to the line emission: in Fig.
\ref{f:intcompMF}, the percentage differences of the line emission, with respect
to the ones obtained using LB98 collisional data, are plotted for different
densities. Once again, the deviations on the emissivities of the H$n\alpha$
lines are linked to the masing range. At densities
$n_\text{H}=10^6 - 10^8 \text{cm}^{-3}$, the differences peak at the frequency
where the optical depths change sign from negative to positive. In both cases,
for frequencies higher than this point ($\sim20$ GHz for
$n_\text{H}=10^6 \text{cm}^{-3}$ and $\sim200$ GHz for
$n_\text{H}=10^8 \text{cm}^{-3}$), the higher collisionality of VS80 and VR62
overcompensates the slightly less negative $\tau$, populating the higher levels
and producing up to 10\% more intense H$n\alpha$ lines. However, those
simulations also have a change of the sign of the optical depths at slightly
higher frequencies than the ones using LB98 or Fujimoto78 data. At this point
($\nu\sim300$ GHz for VS80), the H$n\alpha$ lines are still mased for the other
datasets (LB98 and Fujimoto78), so they predict higher intensities, up to 20\%
for $n_\text{H}=10^8 \text{cm}^{-3}$ and up to 10\% for
$n_\text{H}=10^6 \text{cm}^{-3}$ respectively, for the range of frequencies
down to the frequency point where they have an opacity similar to the other
simulations.

For the high densities represented in Figs. \ref{f:agnODcompMF} and
\ref{f:intcompMF}, the main differences are correlated with the minima of the
optical depths, where the level populations are far away from LTE, and the
uncertainties reach 20\%. At lower frequencies, the curves are parallel and the
intensity differences, as the optical depths, are driven by the differences in
the values of the effective collisional excitation rates. 

\subsection{Orion Model}
\label{sec:orion}

Our model of the Orion Blister consists of an expanding nebula where the
abundance of the species are a mean of the abundances of the Orion Nebula
determined by \citet{Baldwin1991,Rubin1991,Osterbrock1992} and
\citet{Rubin1993}. The large R-size grains distribution described by
\citet{Baldwin1991} is used, but with the physics updated as described by
\citet{VanHoof2004}. Abundances of some rare species were taken from the ISM mix
of \citet{Savage1996}. The ionizing radiation from the O6 star $\theta^1$
Orionis C is simulated from the atmospheres computed by \citet{Castelli2004}.
We assumed a homogeneous density of $\sim 10^4 \text{cm}^{-3}$ and constant
pressure. We have doubled the calculated optical depths in order to account
for the molecular cloud beyond the ionization front, which is not simulated.
This approximation might look rough, but is not important for the purpose
of comparing different collisional rates. Finally the ionizing cosmic rays
background is given by \citet{Indriolo2007}. 

\begin{figure}
  \begin{center}
    \includegraphics[width=0.5\textwidth,clip]{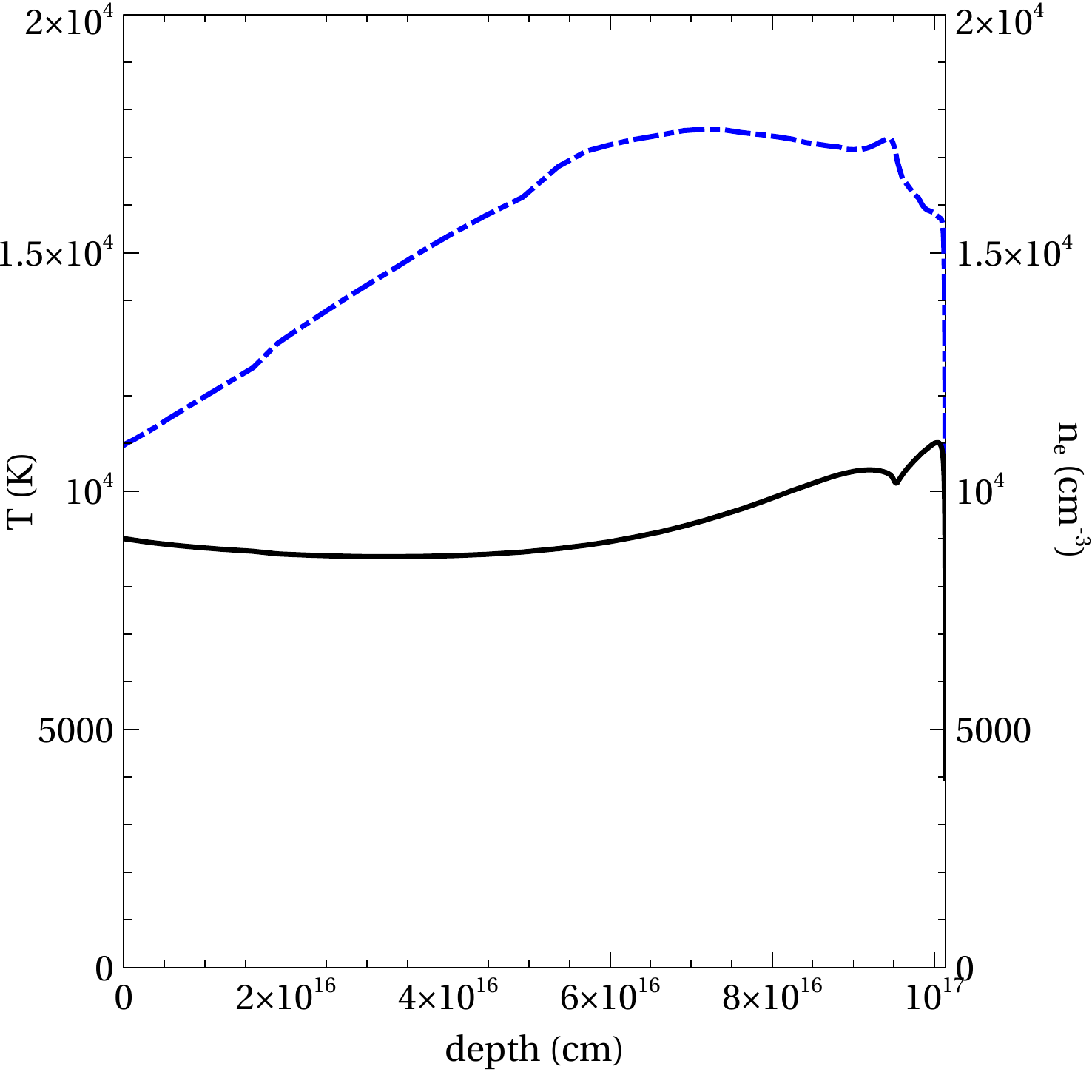}
    \caption{\label{f:temp} Electron temperature (solid black) and electron
      density (dot-dashed blue) in our Orion Blister model as a function of the
      depth of the cloud. }
  \end{center}
\end{figure}

To properly calculate the lines and optical depths of Rydberg levels
transitions, we have extended the hydrogen and helium atoms up to $n=400$. The
levels up to $n=25$ of H-like hydrogen and up to $n=20$ of He-like helium are
resolved in $l$-subshells, while the $l$ populations of the remaining levels
up to $n=400$ are taken as statistically populated. The critical densities for
$l$-changing collisions of $n=20$ and $n=25$ are shown in Table
\ref{t:critdens} for H-like and He-like hydrogen, helium and carbon. The
calculated electron density ranges between $1.09-1.75\times10^4\text{cm}^{-3}$
and is shown in Fig. \ref{f:temp}. Higher nuclear charge ions, such as carbon
ions, have much higher critical densities and are not correctly described, as
seen in Table \ref{t:critdens}. Including more resolved levels for these ions
implies more memory and more computational time, which would render the
calculation impractical. However, note that the main carbon ionization stages
are C$^{2+}$ (Be-like) in the inner part of the nebula and C$^+$ in the outer
parts, and H/He-like C is negligible in any of the zones. If temperature and
radiation conditions would allow for H/He-like ionized carbon, resolved
calculation would be needed for an accurate calculation of the lower levels
recombination lines.

\begin{table}
  \begin{center}
    \caption{\label{t:critdens} Critical densities for $l$-changing processes 
      for $n=20$ and $n=25$ in our Orion nebula model.}
    \begin{tabular}{|c|c|c|}
      \hline
      Ion/Atom & \multicolumn{2}{|c|}{Critical density (cm$^{-3}$ )} \\
      & $n=20$ & $n=25$ \\
      \hline
      H$^0$ & 383 & 54 \\
      \hline
      He$^0$ & 194 & 23 \\
      \hline
      He$^+$ & 18903 & 2605 \\
      \hline
      C$^{4+}$ & $3.136\times10^6$ & $3.711\times10^5$ \\
      \hline
      C$^{5+}$ & $1.911 \times 10^7$ & $2.329\times10^6$ \\
      \hline
    \end{tabular}
  \end{center}
\end{table}

\begin{figure}
  \begin{center}
    \includegraphics[width=0.5\textwidth,clip]{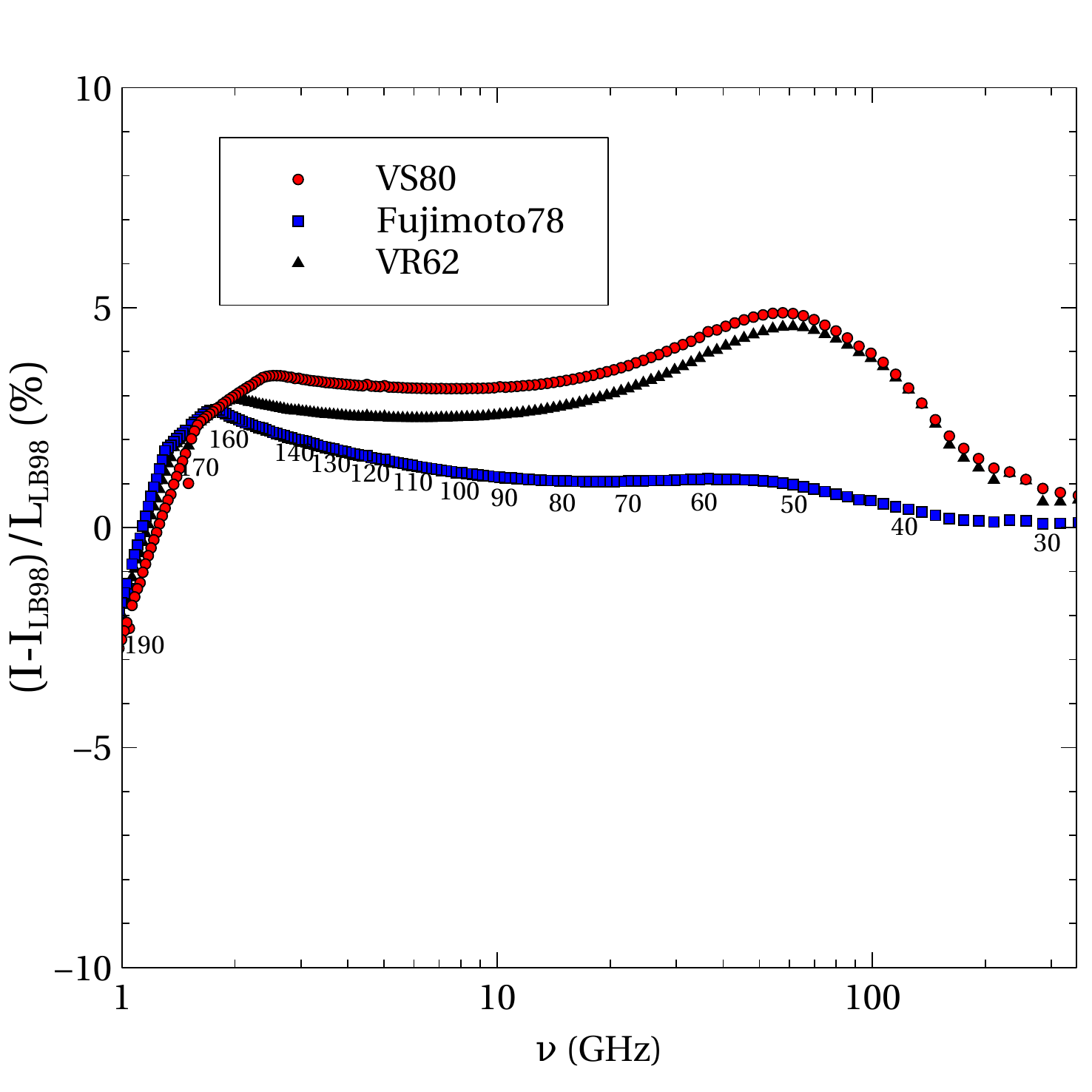}
    \caption{\label{f:alphaemission} Percentage of the ratios of emitted
      H$n\alpha$ transitions with respect to LB98 $n$-changing dataset
      calculations. The small numbers in the figure represent the corresponding
      $n$ of the H$n\alpha$ transition. Datasets of the legend correspond to
      the ones given in Fig. \ref{f:nnm1}.
    }
  \end{center}
\end{figure}

\begin{figure}
  \begin{center}
    \includegraphics[width=0.5\textwidth,clip]{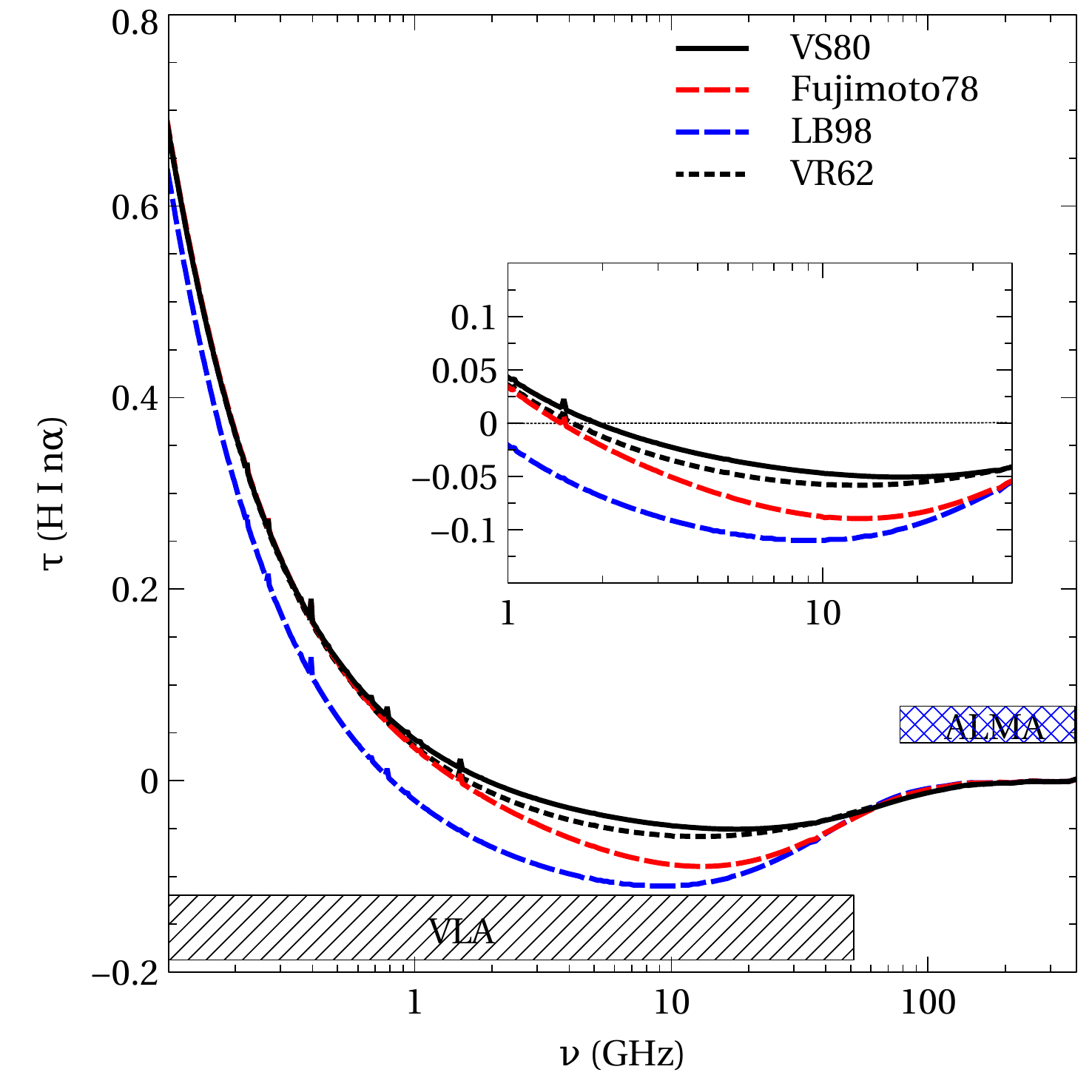}
    \caption{\label{f:optdepth} Optical depths at the wavelengths of the
      H$n\alpha$ lines. The different symbols correspond to the different
      datasets in Fig. \ref{f:nnm1}. The inset is a zoom of the 1-40 GHz zone
      where masing is present. The spikes correspond to the overlapping of
      different lines. The strongest are:at 1.53GHz H$163\alpha$,
      He I $163\alpha$ and H$205\beta$; at 0.79GHz H$203\alpha$ and
      H$292\beta$; at 0.40GHz H$255\alpha$, H$321\beta$ and H$367\gamma$. The
      boxes indicate the frequency range of observation of the ALMA and VLA
      radio telescopes.}
  \end{center}
\end{figure}

In Fig. \ref{f:alphaemission}, the intensities emitted at the surface of the
cloud for the wavelengths of the H$n\alpha$ recombination lines are given for
calculations using the different datasets presented in Section
\ref{sec:datasets}. The percentage difference of the intensities to the ones 
obtained using the LB98 rate coefficients is given as a function of the
frequency. As seen in the Figure, VS80 and VR62 datasets produce a flux that
disagrees with LB98 and Fujimoto78 by 5\% between 2 GHz and 300 GHZ. This is
consistent with the results of Fig. \ref{f:nnm1}. As expected, all datasets
tend to produce similar intensities at frequencies around 1GHz when the system
approaches LTE. The shape of the curves of Fig. \ref{f:alphaemission} can be
explained with the optical depths, shown in Fig. \ref{f:optdepth}, at the
wavelengths of the H$n\alpha$ lines for the different $n$-changing datasets.
The optical depths are negligible for high frequencies down to $\sim40$GHz,
where maser emission starts to take place. The temperature in the ionized
nebula, shown in Fig. \ref{f:temp}, is roughly constant around an average
temperature of 8800K. It corresponds to the zone in Fig. \ref{f:nnm1vst} where
both Fujimoto78 and LB98 agree. From Figs. \ref{f:nnm1} and \ref{f:temp}, in
an optical thin plasma (which is the case here with $\tau\sim0.1$), it is
expected that the difference of intensities stays constant, as the curves in
Fig. \ref{f:nnm1} are parallel for higher $n$ ($n\sim50$). However, the
progressively less negative $\tau$ of VS80 and VR62 and the sustained value of
$\tau$ corresponding to the LB98 dataset, due to the maser effect, partially
cancels the difference. At frequencies lower than 2GHz, the finite optical
depth, that is starting to grow for all datasets, equalizes the fluxes.

The modeled optical depths of Fig. \ref{f:optdepth} are consistent with what is
expected from Fig. \ref{f:betas} and eq. (\ref{eq:kappa}) (note that the
quantities plotted in Fig. \ref{f:betas} are $\log(1-\beta_n)$ ). Even though
the model in Fig. \ref{f:betas} is a simplification, the Orion Blister can be
approximated as a slab of gas and, if the interaction between H ions and the
other species is neglected (charge exchange cross sections at these
temperatures are much lower than radiative and three body recombination cross
sections and the abundances of metals are low \footnote{C and O ions will have
  an important effect in cooling, however they will not affect the abundances of
  H.}), our model in Section \ref{sec:1layer} can be seen as a highly
simplified optically thin version of the one presented in this Section. In
Section \ref{sec:1layer}, all datasets have a negative value of $\beta_n$ up
to $n\sim200$, implying inversion of the populations and the masing of the
H$n\alpha$ lines at these levels (in Fig. \ref{f:optdepth} we obtain negative
$\tau$ of $\sim -0.05 \to -0.1$ ). Higher collisional excitation rates, such
as those of VS80 and VR62, suppress the masing effect by bringing the
population of the Rydberg levels closer to LTE. At low frequencies (high $n$),
when the populations are close to equilibrium, the optical depths converge to
the LTE values. The intensities obtained in our calculations and shown in Figs.
\ref{f:alphaemission} disagree by less than a 5\% and they are within the
observational uncertainties. 

\section{Conclusions}

In this paper we have presented a comparative study of the effect that the 
uncertainties in high $n$ collisional data have on the recombination lines of
Rydberg atoms and how they affect spectroscopic-derived quantities as line
intensities and optical depths. 

Of the different datasets analyzed here and that are widely used by the
astronomical community, two, VS80 \citep{Vriens1980} and Fujimoto78
\citep{Fujimoto1978}, are based on semi-empirical calculations, one, VR62 
\citep{VanRegemorter1962}, is the $\bar{g}$ approximation where the Gaunt
factors have been fitted to experimental results previous to 1962, and the
last one, LB98 \citep{Lebedevandbeigman1998}, are \emph{ab initio} impact 
parameter Born approximation results. All datasets presented here show a
reasonable level of agreement, to within a factor $\sim$2. However, our
simulations predict unexpectedly different emission in H$n\alpha$ RRL for some
frequency ranges. This is related to the effectiveness of the electron
(de-)excitation collisions in the redistribution of the high Rydberg electron
population. This changes the level where the H$n\alpha$ lines stop masing, i.e.
the point where the optical depth at the H$n\alpha$ frequencies changes sign
from negative to positive.

In order to check the importance of the collisional excitation uncertainties in 
the modeling of astronomical objects, we have modeled the RRL emission in BLRs 
and the Orion Blister H~II region. \citet{Manti2016,Manti2017} postulate that
RRL from obscured QSO could be easily detectable with the SKA radio-telescope,
while \citet{Izumi2016} predict that sub-mm RRL would be too faint to be
observed, even with ALMA. However, we have noticed that the latter estimation
was done assuming that levels with principal quantum number $n\simeq20$ would
be close to LTE, something that does not happen in our simulations. BLRs
physical conditions, and thus its emission, can depend on the energy input
(SED) and the density. Differences in the assumed SED changes the length of
the ionized region and can produce changes on the emission of H$n\alpha$.
Different densities can determine where the H$n\alpha$ lines can be mased and
produce a systematic offset in the absolute intensity emitted by the cloud.
The error introduced by the uncertainty in the collisional data is constrained
to a specific frequency range, usually between 40 GHz and 100 GHz. Differences
on emission H$n\alpha$ lines can amount up to 20\% for
$n_\text{H} \geq 10^8 \text{cm}^{-3}$ and 10\% for
$n_\text{H} = 10^6 \text{cm}^{-3}$. The Orion Blister has been simulated using
the model described by \citet{Baldwin1991}. We see differences in the optical
depths that are reflected in uncertainties of $\sim5$\% in the H$n\alpha$ RRL
which are not relevant for astronomical observations.

Uncertainties of $\sim60$\% on the electron impact excitation data propagate to
$\sim5$\% error on the predicted emission intensities. We do not expect that
these differences could be higher than the experimental error. However, in
situations where natural masing of the Rydberg levels are expected, these
differences can increase up to a significant 20\% as has been shown here for
BLRs. We note that, when neutral gas resides beyond the ionization front of
BLRs, $n$-changing collisions could be relevant in the transition zone where
the H and He are recombining and produce a significant effect on the emitted
intensities. Future work will clarify this point.

Accurate collisional data might play a significant role in the determination 
of the electron fraction and the radiation during the recombination era.
\citet{Chlubaandalihaimoud2016} (in their recombination code COSMOSPEC) use 
the collisional rates from \cite{VanRegemorter1962}, while in other 
calculations \citep[e.g][]{GrinandHirata2010,alihaimoudandhirata2011} 
collisions are neglected. \citet{Chlubaandalihaimoud2016} report small effects 
on the recombination spectrum at low frequencies after increasing 
(de-)excitation collisions two orders of magnitude. However, they point out 
the necessity of including collisional data for precise computations of the 
recombination radiation. Collisional data could be important especially at low 
redshift ($z\lesssim 1000$) when matter is more diluted and the differences 
between the collisional rates might be seen in the free electron fraction. 
$n$-changing collisions bring the higher levels to LTE and suppress the 
emission of photons from decaying electrons. In addition, population inversion 
can exist for highly excited levels during the recombination and reonization 
epochs resulting in local masing in dense forming structures 
\citep{GrinandHirata2010}. These masers could also be strongly affected by 
energy changing collisions.

\section{Acknowledgments}

The authors acknowledge support from the National Science Foundation
(AST-1816537) and NASA ATP program (grant number 17-ATP17-0141). MC
acknowledges support from STScI (HST-AR-14286.001-A and HST-AR-14556.001-A). PvH
acknowledges support from the Belgian Science Policy Office through  contract
no. BR/143/A2/BRASS. NRB acknowledges support from STFC (UK) through the
University of Strathclyde APAP Network grant ST/R000743/1. M.D.\ and 
G.F.\ acknowledge support from the NSF (AST-1816537), NASA (ATP 17-0141),
and STScI (HST-AR-13914, HST-AR-15018). The National Radio Astronomy
Observatory is a Facility of the Nacional Science Foundation operated under
cooperative agreement by Associated University, Inc.

\appendix
 \section{Importance of a complete model}
 \label{sec:atommodel}

\begin{figure}
  \begin{center}
    \includegraphics[width=0.5\textwidth,clip]{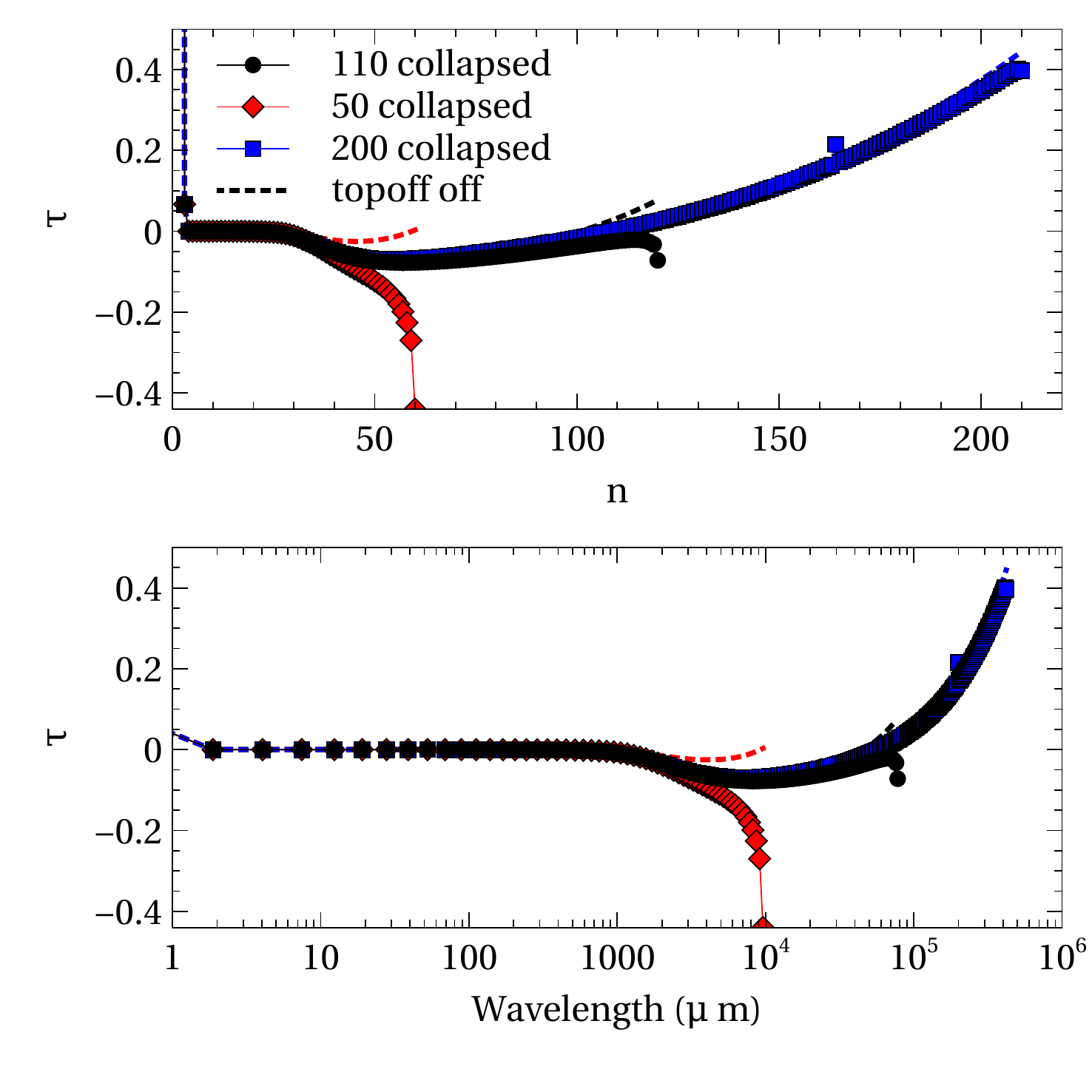}
    \caption{\label{f:numlevels} Variation of the optical depth of the
      Hn$\alpha$ lines for a slab of pure hydrogen gas irradiated by black
      body radiation at $T=10^5$K including a different number of levels. Top:
      Optical depth vs. highest principal quantum number. Bottom: Optical
      depth vs. wavelength.}
  \end{center}
\end{figure}

 Maser effects occur for non-LTE RRL around sub-mm frequencies. This affects
 ALMA measurements \citep{Peters2012}. The simulations of RRL spectra need to
 include a high number of Rydberg levels, and many of them will not be in LTE
 for standard cloud densities (see Fig. \ref{f:bnvsdens}). The Cloudy
 simulations used throughout this paper were done by setting the model atom up
 to high $n$. In an ideal atom, the highest levels (at $n\to\infty$) are in LTE
 with the continuum and neighboring levels. However, an infinite levels atom
 is impractical in models. In Cloudy, we can use a ``top-off'' where the
 highest level considered is forced in LTE with the continuum by enhancing its
 collisional ionization and three body recombination. This is an imperfect
 treatment of the atom and can lead to errors in the populations. These errors
 can be reduced by the fact that at intermediate densities the continuum is
 lowered to finite $n$ by the effect of electron pressure \citep{Bautista00}.
 To illustrate the lack of accuracy in the high $n$ populations, we have used a
 simple model of a slab of pure hydrogen gas at $T=7500$K and
 $n_{\text{H}}=10^5\text{cm}^{-3}$, divided in 100 zones and irradiated by a
 black-body radiation at $T=10^5$K. We have then calculated the optical depth
 for different atom sizes with a different number of collapsed levels. The
 results for the optical depth are in Fig. \ref{f:numlevels}. When the number
 of levels is not big enough, the enhanced population of the top level
 produces an increment of the population inversion and an artificial increase
 of the line emission. If the ``top-off'' is disabled, the effect is the
 contrary as the population of the top level does not account for the cascade
 from higher levels. The real solution dwells between these extremes. When
 more and more levels are included in the calculation, the ionization and
 three body recombination coefficients bring the top level closer to LTE and
 the effect of the ``top-off'' mechanism is diluted.

\bibliographystyle{mn2e}
\bibliography{nchanging_final.bbl}

\begin{thebibliography}{}

\bibitem[\protect\citeauthoryear{{Abramowitz} \& {Stegun}}{{Abramowitz} \&
  {Stegun}}{1965}]{abramowitz1965}
{Abramowitz} M.,  {Stegun} I.~A.,  1965, {Handbook of mathematical functions
  with formulas, graphs, and mathematical tables, Corrected edition, edited by
  Abramowitz, Milton; Stegun, Irene A.}.
Dover Books on Advanced Mathematics, New York: Dover

\bibitem[\protect\citeauthoryear{{Ali-Ha{\"\i}moud} \&
  {Hirata}}{{Ali-Ha{\"\i}moud} \& {Hirata}}{2011}]{alihaimoudandhirata2011}
{Ali-Ha{\"\i}moud} Y.,  {Hirata} C.~M.,  2011, \prd, 83, 043513

\bibitem[\protect\citeauthoryear{Anderson, Ballance, Badnell \&
  Summers}{Anderson et~al.}{2000}]{Anderson2000}
Anderson H.,  Ballance C.~P.,  Badnell N.~R.,    Summers H.~P.,  2000, Journal
  of Physics B: Atomic, Molecular and Optical Physics, 33, 1255

\bibitem[\protect\citeauthoryear{{Anderson}, {Armentrout}, {Johnstone},
  {Bania}, {Balser}, {Wenger} \& {Cunningham}}{{Anderson}
  et~al.}{2015}]{Anderson2015}
{Anderson} L.~D.,  {Armentrout} W.~P.,  {Johnstone} B.~M.,  {Bania} T.~M.,
  {Balser} D.~S.,  {Wenger} T.~V.,    {Cunningham} V.,  2015, \apjs, 221, 26

\bibitem[\protect\citeauthoryear{{Anderson}, {Armentrout}, {Luisi}, {Bania},
  {Balser} \& {Wenger}}{{Anderson} et~al.}{2018}]{Anderson2018}
{Anderson} L.~D.,  {Armentrout} W.~P.,  {Luisi} M.,  {Bania} T.~M.,  {Balser}
  D.~S.,    {Wenger} T.~V.,  2018, The Astrophysical Journal Supplement Series,
  234, 33

\bibitem[\protect\citeauthoryear{{Baker} \& {Menzel}}{{Baker} \&
  {Menzel}}{1938}]{Baker1938}
{Baker} J.~G.,  {Menzel} D.~H.,  1938, \apj, 88, 52

\bibitem[\protect\citeauthoryear{{Baldwin}, {Ferland}, {Martin}, {Corbin},
  {Cota}, {Peterson} \& {Slettebak}}{{Baldwin} et~al.}{1991}]{Baldwin1991}
{Baldwin} J.~A.,  {Ferland} G.~J.,  {Martin} P.~G.,  {Corbin} M.~R.,  {Cota}
  S.~A.,  {Peterson} B.~M.,    {Slettebak} A.,  1991, \apj, 374, 580

\bibitem[\protect\citeauthoryear{{Ballance}, {Griffin}, {Loch}, {Boivin} \&
  {Pindzola}}{{Ballance} et~al.}{2006}]{Ballance2006}
{Ballance} C.~P.,  {Griffin} D.~C.,  {Loch} S.~D.,  {Boivin} R.~F.,
  {Pindzola} M.~S.,  2006, \pra, 74, 012719

\bibitem[\protect\citeauthoryear{{Balser}, {Rood}, {Bania} \&
  {Anderson}}{{Balser} et~al.}{2011}]{Balser2011}
{Balser} D.~S.,  {Rood} R.~T.,  {Bania} T.~M.,    {Anderson} L.~D.,  2011,
  \apj, 738, 27

\bibitem[\protect\citeauthoryear{{Balser}, {Wenger}, {Anderson} \&
  {Bania}}{{Balser} et~al.}{2015}]{Balser2015}
{Balser} D.~S.,  {Wenger} T.~V.,  {Anderson} L.~D.,    {Bania} T.~M.,  2015,
  \apj, 806, 199

\bibitem[\protect\citeauthoryear{{Bautista} \& {Kallman}}{{Bautista} \&
  {Kallman}}{2000}]{Bautista00}
{Bautista} M.~A.,  {Kallman} T.~R.,  2000, \apj, 544, 581

\bibitem[\protect\citeauthoryear{{Bray}, {Burgess}, {Fursa} \& {Tully}}{{Bray}
  et~al.}{2000}]{Bray2000}
{Bray} I.,  {Burgess} A.,  {Fursa} D.~V.,    {Tully} J.~A.,  2000, \aaps, 146,
  481

\bibitem[\protect\citeauthoryear{{Brocklehurst}}{{Brocklehurst}}{1970a}]{Brocklehurst1970}
{Brocklehurst} M.,  1970a, \mnras, 148, 417

\bibitem[\protect\citeauthoryear{{Brocklehurst}}{{Brocklehurst}}{1970b}]{Brocklehurst.Nat.1970}
{Brocklehurst} M.,  1970b, \nat, 225, 618

\bibitem[\protect\citeauthoryear{{Brocklehurst} \& {Salem}}{{Brocklehurst} \&
  {Salem}}{1977}]{Brocklehurstandsalem1977}
{Brocklehurst} M.,  {Salem} M.,  1977, Computer Physics Communications, 13, 39

\bibitem[\protect\citeauthoryear{{Brocklehurst} \& {Seaton}}{{Brocklehurst} \&
  {Seaton}}{1972}]{Brocklehurstandseaton1972}
{Brocklehurst} M.,  {Seaton} M.~J.,  1972, \mnras, 157, 179

\bibitem[\protect\citeauthoryear{{Brown}, {Lockman} \& {Knapp}}{{Brown}
  et~al.}{1978}]{Brown1978}
{Brown} R.~L.,  {Lockman} F.~J.,    {Knapp} G.~R.,  1978, \araa, 16, 445

\bibitem[\protect\citeauthoryear{{Burgess} \& {Summers}}{{Burgess} \&
  {Summers}}{1976}]{Burgess1976}
{Burgess} A.,  {Summers} H.~P.,  1976, \mnras, 174, 345

\bibitem[\protect\citeauthoryear{{Castelli} \& {Kurucz}}{{Castelli} \&
  {Kurucz}}{2004}]{Castelli2004}
{Castelli} F.,  {Kurucz} R.~L.,  2004, arXiv:astro-ph/0405087

\bibitem[\protect\citeauthoryear{{Chluba} \& {Ali-Ha{\"\i}moud}}{{Chluba} \&
  {Ali-Ha{\"\i}moud}}{2016}]{Chlubaandalihaimoud2016}
{Chluba} J.,  {Ali-Ha{\"\i}moud} Y.,  2016, \mnras, 456, 3494

\bibitem[\protect\citeauthoryear{{Dupree} \& {Goldberg}}{{Dupree} \&
  {Goldberg}}{1970}]{Dupreeandgoldberg1970}
{Dupree} A.~K.,  {Goldberg} L.,  1970, \araa, 8, 231

\bibitem[\protect\citeauthoryear{{Ferland}, {Chatzikos}, {Guzm{\'a}n},
  {Lykins}, {van Hoof}, {Williams}, {Abel}, {Badnell}, {Keenan}, {Porter} \&
  {Stancil}}{{Ferland} et~al.}{2017}]{CloudyReview}
{Ferland} G.~J.,  {Chatzikos} M.,  {Guzm{\'a}n} F.,  {Lykins} M.~L.,  {van
  Hoof} P.~A.~M.,  {Williams} R.~J.~R.,  {Abel} N.~P.,  {Badnell} N.~R.,
  {Keenan} F.~P.,  {Porter} R.~L.,    {Stancil} P.~C.,  2017, Revista Mexicana
  de Astronomia y Astrofisica, 53, 385

\bibitem[\protect\citeauthoryear{Fujimoto}{Fujimoto}{1978}]{Fujimoto1978}
Fujimoto T.,  1978, Technical Report IPPJ-AM-8, Semi-empirical Cross Sections
  and Rate Coefficients for Excitation and Ionization by Electron Collision and
  Photoionization of Helium.
Nagoya University

\bibitem[\protect\citeauthoryear{{Gee}, {Percival}, {Lodge} \&
  {Richards}}{{Gee} et~al.}{1976}]{Gee1976}
{Gee} C.~S.,  {Percival} L.~C.,  {Lodge} J.~G.,    {Richards} D.,  1976,
  \mnras, 175, 209

\bibitem[\protect\citeauthoryear{{Goldberg}}{{Goldberg}}{1966}]{Goldberg1966}
{Goldberg} L.,  1966, \apj, 144, 1225

\bibitem[\protect\citeauthoryear{{Griffin} \& {Ballance}}{{Griffin} \&
  {Ballance}}{2009}]{Griffinandballance2009}
{Griffin} D.~C.,  {Ballance} C.~P.,  2009, Journal of Physics B Atomic
  Molecular Physics, 42, 235201

\bibitem[\protect\citeauthoryear{{Grin} \& {Hirata}}{{Grin} \&
  {Hirata}}{2010}]{GrinandHirata2010}
{Grin} D.,  {Hirata} C.~M.,  2010, \prd, 81, 083005

\bibitem[\protect\citeauthoryear{{Guzm{\'a}n}, {Badnell}, {Williams}, {van
  Hoof}, {Chatzikos} \& {Ferland}}{{Guzm{\'a}n} et~al.}{2016}]{Guzman.I.2016}
{Guzm{\'a}n} F.,  {Badnell} N.~R.,  {Williams} R.~J.~R.,  {van Hoof} P.~A.~M.,
  {Chatzikos} M.,    {Ferland} G.~J.,  2016, \mnras, 459, 3498

\bibitem[\protect\citeauthoryear{{Guzm{\'a}n}, {Badnell}, {Williams}, {van
  Hoof}, {Chatzikos} \& {Ferland}}{{Guzm{\'a}n} et~al.}{2017}]{Guzman.II.2017}
{Guzm{\'a}n} F.,  {Badnell} N.~R.,  {Williams} R.~J.~R.,  {van Hoof} P.~A.~M.,
  {Chatzikos} M.,    {Ferland} G.~J.,  2017, \mnras, 464, 312

\bibitem[\protect\citeauthoryear{{Indriolo}, {Geballe}, {Oka} \&
  {McCall}}{{Indriolo} et~al.}{2007}]{Indriolo2007}
{Indriolo} N.,  {Geballe} T.~R.,  {Oka} T.,    {McCall} B.~J.,  2007, \apj,
  671, 1736

\bibitem[\protect\citeauthoryear{Izotov, Thuan \& Guseva}{Izotov
  et~al.}{2014}]{Izotov2014}
Izotov Y.~I.,  Thuan T.~X.,    Guseva N.~G.,  2014, Monthly Notices of the
  Royal Astronomical Society, 445, 778

\bibitem[\protect\citeauthoryear{{Izotov}, {Thuan} \& {Lipovetsky}}{{Izotov}
  et~al.}{1997}]{Izotov1997}
{Izotov} Y.~I.,  {Thuan} T.~X.,    {Lipovetsky} V.~A.,  1997, The Astrophysical
  Journal Supplement Series, 108, 1

\bibitem[\protect\citeauthoryear{{Izumi}, {Nakanishi}, {Imanishi} \&
  {Kohno}}{{Izumi} et~al.}{2016}]{Izumi2016}
{Izumi} T.,  {Nakanishi} K.,  {Imanishi} M.,    {Kohno} K.,  2016, \mnras, 459,
  3629

\bibitem[\protect\citeauthoryear{{Johnson}}{{Johnson}}{1972}]{Johnson1972}
{Johnson} L.~C.,  1972, \apj, 174, 227

\bibitem[\protect\citeauthoryear{{Johnson} \& {Hinnov}}{{Johnson} \&
  {Hinnov}}{1969}]{Johnsonandhinnov1969}
{Johnson} L.~C.,  {Hinnov} E.,  1969, Physical Review, 187, 143

\bibitem[\protect\citeauthoryear{{Korista}, {Baldwin}, {Ferland} \&
  {Verner}}{{Korista} et~al.}{1997}]{KoristaBaldwin1997}
{Korista} K.,  {Baldwin} J.,  {Ferland} G.,    {Verner} D.,  1997, \apjs, 108,
  401

\bibitem[\protect\citeauthoryear{{Lebedev} \& {Beigman}}{{Lebedev} \&
  {Beigman}}{1998}]{Lebedevandbeigman1998}
{Lebedev} V.~S.,  {Beigman} I.~L.,  1998, {Physics of Highly Excited Atoms and
  Ions}.
Springer

\bibitem[\protect\citeauthoryear{{Luisi}, {Anderson}, {Balser}, {Wenger} \&
  {Bania}}{{Luisi} et~al.}{2017}]{Luisi2017}
{Luisi} M.,  {Anderson} L.~D.,  {Balser} D.~S.,  {Wenger} T.~V.,    {Bania}
  T.~M.,  2017, \apj, 849, 117

\bibitem[\protect\citeauthoryear{{Manti}, {Gallerani}, {Ferrara}, {Feruglio},
  {Graziani} \& {Bernardi}}{{Manti} et~al.}{2016}]{Manti2016}
{Manti} S.,  {Gallerani} S.,  {Ferrara} A.,  {Feruglio} C.,  {Graziani} L.,
  {Bernardi} G.,  2016, \mnras, 456, 98

\bibitem[\protect\citeauthoryear{{Manti}, {Gallerani}, {Ferrara}, {Greig} \&
  {Feruglio}}{{Manti} et~al.}{2017}]{Manti2017}
{Manti} S.,  {Gallerani} S.,  {Ferrara} A.,  {Greig} B.,    {Feruglio} C.,
  2017, \mnras, 466, 1160

\bibitem[\protect\citeauthoryear{{Mathews} \& {Ferland}}{{Mathews} \&
  {Ferland}}{1987}]{Mathews1987}
{Mathews} W.~G.,  {Ferland} G.~J.,  1987, \apj, 323, 456

\bibitem[\protect\citeauthoryear{{Morabito}, {Oonk}, {Salgado}, {Toribio},
  {R{\"o}ttgering}, {Tielens}, {Beck}, {Adebahr}, {Best}, {Beswick} \& et
  al}{{Morabito} et~al.}{2014}]{Morabito2014}
{Morabito} L.~K.,  {Oonk} J.~B.~R.,  {Salgado} F.,  {Toribio} M.~C.,
  {R{\"o}ttgering} H.~J.~A.,  {Tielens} A.~G.~G.~M.,  {Beck} R.,  {Adebahr} B.,
   {Best} P.,  {Beswick} R.,    et al 2014, \apjl, 795, L33

\bibitem[\protect\citeauthoryear{{Olive}, {Steigman} \& {Walker}}{{Olive}
  et~al.}{2000}]{Olive2000}
{Olive} K.~A.,  {Steigman} G.,    {Walker} T.~P.,  2000, \physrep, 333, 389

\bibitem[\protect\citeauthoryear{{Osterbrock} \& {Ferland}}{{Osterbrock} \&
  {Ferland}}{2006}]{AGN3}
{Osterbrock} D.~E.,  {Ferland} G.~J.,  2006, {Astrophysics of gaseous nebulae
  and active galactic nuclei, 2nd.~ed.}.
Sausalito, CA: University Science Books

\bibitem[\protect\citeauthoryear{{Osterbrock}, {Tran} \&
  {Veilleux}}{{Osterbrock} et~al.}{1992}]{Osterbrock1992}
{Osterbrock} D.~E.,  {Tran} H.~D.,    {Veilleux} S.,  1992, \apj, 389, 305

\bibitem[\protect\citeauthoryear{{Pengelly} \& {Seaton}}{{Pengelly} \&
  {Seaton}}{1964}]{PengellySeaton1964}
{Pengelly} R.~M.,  {Seaton} M.~J.,  1964, \mnras, 127, 165

\bibitem[\protect\citeauthoryear{{Percival} \& {Richards}}{{Percival} \&
  {Richards}}{1976}]{Percival1976}
{Percival} I.~C.,  {Richards} D.,  1976, Advances in Atomic and Molecular
  Physics, 11, 1

\bibitem[\protect\citeauthoryear{{Percival} \& {Richards}}{{Percival} \&
  {Richards}}{1978}]{Percival1978}
{Percival} I.~C.,  {Richards} D.,  1978, \mnras, 183, 329

\bibitem[\protect\citeauthoryear{{Peters}, {Longmore} \& {Dullemond}}{{Peters}
  et~al.}{2012}]{Peters2012}
{Peters} T.,  {Longmore} S.~N.,    {Dullemond} C.~P.,  2012, \mnras, 425, 2352

\bibitem[\protect\citeauthoryear{{Poppi}, {Tsivilev}, {Cortiglioni}, {Palumbo}
  \& {Sorochenko}}{{Poppi} et~al.}{2007}]{Poppi2007}
{Poppi} S.,  {Tsivilev} A.~P.,  {Cortiglioni} S.,  {Palumbo} G.~G.~C.,
  {Sorochenko} R.~L.,  2007, \aap, 464, 995

\bibitem[\protect\citeauthoryear{{Ralchenko}, {Janev}, {Kato}, {Fursa}, {Bray}
  \& {de Heer}}{{Ralchenko} et~al.}{2008}]{Ralchenko2008}
{Ralchenko} Y.,  {Janev} R.~K.,  {Kato} T.,  {Fursa} D.~V.,  {Bray} I.,    {de
  Heer} F.~J.,  2008, Atomic Data and Nuclear Data Tables, 94, 603

\bibitem[\protect\citeauthoryear{{Rohlfs} \& {Wilson}}{{Rohlfs} \&
  {Wilson}}{2000}]{Rohlfsandwilson2000}
{Rohlfs} K.,  {Wilson} T.~L.,  2000, {Tools of radio astronomy}.
Springer

\bibitem[\protect\citeauthoryear{{Rubin}, {Dufour} \& {Walter}}{{Rubin}
  et~al.}{1993}]{Rubin1993}
{Rubin} R.~H.,  {Dufour} R.~J.,    {Walter} D.~K.,  1993, \apj, 413, 242

\bibitem[\protect\citeauthoryear{{Rubin}, {Simpson}, {Haas} \&
  {Erickson}}{{Rubin} et~al.}{1991}]{Rubin1991}
{Rubin} R.~H.,  {Simpson} J.~P.,  {Haas} M.~R.,    {Erickson} E.~F.,  1991,
  \apj, 374, 564

\bibitem[\protect\citeauthoryear{{Salgado}, {Morabito}, {Oonk}, {Salas},
  {Toribio}, {R{\"o}ttgering} \& {Tielens}}{{Salgado}
  et~al.}{2017}]{Salgado.I.2017}
{Salgado} F.,  {Morabito} L.~K.,  {Oonk} J.~B.~R.,  {Salas} P.,  {Toribio}
  M.~C.,  {R{\"o}ttgering} H.~J.~A.,    {Tielens} A.~G.~G.~M.,  2017, \apj,
  837, 141

\bibitem[\protect\citeauthoryear{{Savage} \& {Sembach}}{{Savage} \&
  {Sembach}}{1996}]{Savage1996}
{Savage} B.~D.,  {Sembach} K.~R.,  1996, \araa, 34, 279

\bibitem[\protect\citeauthoryear{{Scoville} \& {Murchikova}}{{Scoville} \&
  {Murchikova}}{2013}]{Scoville2013}
{Scoville} N.,  {Murchikova} L.,  2013, \apj, 779, 75

\bibitem[\protect\citeauthoryear{{Seaton}}{{Seaton}}{1959}]{Seaton1959a}
{Seaton} M.~J.,  1959, \mnras, 119, 81

\bibitem[\protect\citeauthoryear{{Seaton}}{{Seaton}}{1962}]{Seaton1962}
{Seaton} M.~J.,  1962, Proceedings of the Physical Society, 79, 1105

\bibitem[\protect\citeauthoryear{{Shaver}}{{Shaver}}{1980}]{Shaver1980}
{Shaver} P.~A.,  1980, \aap, 91, 279

\bibitem[\protect\citeauthoryear{Sinclair, Yoshikawa, Harries, Young, Weimer \&
  Johnson}{Sinclair et~al.}{1965}]{Sinclair1965}
Sinclair R.~M.,  Yoshikawa S.,  Harries W.~L.,  Young K.~M.,  Weimer K.~E.,
  Johnson J.~L.,  1965, The Physics of Fluids, 8, 118

\bibitem[\protect\citeauthoryear{{Storey} \& {Hummer}}{{Storey} \&
  {Hummer}}{1995}]{Storey1995}
{Storey} P.~J.,  {Hummer} D.~G.,  1995, \mnras, 272, 41

\bibitem[\protect\citeauthoryear{{van Hoof}, {Weingartner}, {Martin}, {Volk} \&
  {Ferland}}{{van Hoof} et~al.}{2004}]{VanHoof2004}
{van Hoof} P.~A.~M.,  {Weingartner} J.~C.,  {Martin} P.~G.,  {Volk} K.,
  {Ferland} G.~J.,  2004, \mnras, 350, 1330

\bibitem[\protect\citeauthoryear{{van Regemorter}}{{van
  Regemorter}}{1962}]{VanRegemorter1962}
{van Regemorter} H.,  1962, \apj, 136, 906

\bibitem[\protect\citeauthoryear{{Vriens} \& {Smeets}}{{Vriens} \&
  {Smeets}}{1980}]{Vriens1980}
{Vriens} L.,  {Smeets} A.~H.~M.,  1980, \pra, 22, 940

\end{thebibliography}
\bsp

\label{lastpage}
\clearpage
\end{document}